\documentclass[12pt]{article}
\usepackage{amsmath,amssymb,exscale,multirow,graphicx}
\usepackage[utf8]{inputenc}
\usepackage{xcolor}
\usepackage{cite}
\usepackage{hyperref}

\input epsf
\newcommand{\m}[1]{\marginpar{{\tiny *}} }
\newcommand{\pslash}{{\not \!p}}

\def\det{{\rm det}}

\def\O{\mathcal{O}}

\def\be{\begin{equation}}
\def\te{\end{equation}} 
\def\nn{\nonumber}  
\def\Tr{\textrm{Tr}}

\def\a{\alpha}
\def\b{\beta}
\def\0{{(0)}}

\def\al#1\fal{\begin{align}#1\end{align}} 
 
\def\d{\delta}

\def\tim{\!\times\!}
\def\C{\Pi}
\def\d1{\delta}
\def\da{\downarrow}
\def\ua{\uparrow}
\def\rfttoo{{\bf r}_{l_b}}
\def\rc{\bf(\bar{4},2,3,1,1)}
\def\rs{{\bf r}_{e_b}}
\def\ro{\bf(20,1,2,2,2)}
\def\roa{{\bf r}_q}
\def\roab{{\bf\bar r}_q}
\def\rob{{\bf r}_{l_a}}
\def\robb{{\bf \bar r}_q}
\def\rfa{{\bf r}_{u}}
\def\rfbb{{\bf \bar r}_{e_a}}
\def\rq{{\bf r}_q}
\def\rl{{\bf r}_{l_a}}
\def\ru{{\bf r}_u}
\def\re{{\bf r}_{e_a}}
\def\rlb{{\bf r}_{l_b}}
\def\reb{{\bf r}_{e_b}}

\def\g{g_{*}}
\newcommand\xr[1]{x_{u,#1}}
\newcommand\xt[1]{x_{3,#1}}
\newcommand\xs[1]{x_{1,#1}}
\newcommand\so[1]{{\rm SO}\!\left(#1\right)}
\newcommand\su[1]{{\rm SU}\!\left(#1\right)}
\newcommand\uu[1]{{\rm U}\!\left(#1\right)}

\begin{document}
\topmargin -1.0cm
\oddsidemargin -0.8cm
\evensidemargin -0.8cm

\begin{center}
\vspace{40pt}

\Large \textbf{A model for the Singlet-Triplet Leptoquarks}

\end{center}

\vspace{15pt}
\begin{center}
{\bf Leandro Da Rold$^{\circ}$, Federico Lamagna$^{\star}$} 

\vspace{20pt}

\textit{Centro At\'omico Bariloche, Instituto Balseiro and CONICET}
\\[0.2cm]
\textit{Av.\ Bustillo 9500, 8400, S.\ C.\ de Bariloche, Argentina}

\end{center}

\vspace{20pt}
\begin{center}
\textbf{Abstract}
\end{center}
\vspace{5pt} {\small \noindent
The deviations of $B$-meson decays measured in $R_{D^{(*)}}^{\tau\ell}$ and $R_{K^{(*)}}^{\mu e}$ can be explained by the presence of two scalar leptoquarks, a singlet $S_1$ and a triplet $S_3$, mostly coupled to the third generation. We consider a theory of resonances, as an effective description of a strongly interacting theory, that generates the leptoquarks and the Higgs as Nambu-Goldstone bosons, with the rest of the resonances at a scale of order $10-30\ {\rm TeV}$. We assume anarchic partial compositeness for the flavor of the SM fermions. Under these hypothesis we study whether it is possible to reproduce the deviations in the $B$-decays without being in conflict with flavor and electroweak bounds. We find a tension between $R_{D^{(*)}}^{\tau\ell}$ and some flavor observables, dominated by flavor violating $\tau$ decays and $\Delta m_{B_s}$, that require a tuning of order $10-25\%$. We also compute the potential of the scalars showing that leptoquarks with masses ${\cal}O(2-3)\ {\rm TeV}$ can be naturally expected in the model. We discuss briefly the phenomenology of the other resonances.
}

\vfill
\noindent {\footnotesize E-mail:
$\circ$ daroldl@cab.cnea.gov.ar,
$\star$ federico.lamagna@cab.cnea.gov.ar
}

\noindent
\eject

\tableofcontents
\section{Introduction}
Despite the lack of direct evidence of new physics at LHC, in the last years several deviations from the Standard Model (SM) predictions have been measured in the decay of $B$-mesons, pointing towards the violation of lepton flavor universality (LFU). The deviations, also referred in the literature as anomalies, are observed in charged current processes involving $b\to c\tau\bar\nu$, as well as in neutral current interactions involving $b\to s\mu\bar\mu$. The first ones, that show deviations from $\tau/\ell$ universality slightly larger than $3\sigma$, have been measured in the ratio $R_{D^{(*)}}^{\tau\ell}$, in different experiments~\cite{RD_Babar,RD_Belle,RD_LHCB}. The second ones, that show deviations of order $4\sigma$ from $\mu/e$ universality, have been measured at LHCb and Belle in the ratio $R_{K^{(*)}}^{\mu e}$~\cite{RK_LHCB,RK_LHCB2,RK_LHCB3,RK_Belle}.

Although there is no evidence of a common origin of the deviations, given that both involve LFU violation in $B$-decays, it seems interesting to attempt a common explanation. 
Whereas the deviations in the aforementioned semileptonic $B$-decays are rather large, compared with the SM amplitudes, no effects have been observed in $K$ and $\pi$ decays, as well as in $\tau$ decays. 

The interpretation of the $B$-anomalies as an effect of new physics have triggered a lot of work in the last years.
Under this hypothesis, the deviations could be explained if the new physics interacts mostly with the third generation, with small but non-negligible interactions with the second one, and tiny or zero interactions with the first one. 
This hierarchy of interactions can be naturally obtained if the mediators are resonances of a strongly interacting theory at a few TeV scale, with the SM fermions being partially composite. 
This scenario is similar to modern composite Higgs models, where the new dynamics stabilizing the Higgs potential is mainly coupled to the heavy fermions of the SM. 
Even though there is no reason for a connection between these issues, it seems appealing to study them in a common framework.

There have been several proposals for a common explanation of the $B$-anomalies, from the perspective of effective field theories~\cite{EFT_Azatov,EFT_Bhattacharya,EFT_Grinstein,EFT_Greljo,EFT_Calibbi,EFT_Bordone,Marzocca_1,Marzocca_2,Cata1,Cata2} 
and also with models containing new states and interactions~\cite{NP_Bauer,NP_Fajfer,NP_Barbieri,NP_Das,NP_Crivellin,NP_Becirevic,NP_Boucenna,NP_Hiller,NP_Bhattacharya,NP_Crivellin2,NP_Barbieri2,NP_Cai,NP_Megias,NP_Popov,NP_Angelescu,NP_Cornella,NP_Marzocca,Isidori,Crivellin,NP_Becirevic2,Bhupal_1,Bhupal_2,GUT_Fajfer,Stangl,Marzocca_2,Cata1,Cata2,Megias}. One of the most appealing hypothesis is the presence of leptoquarks (LQs) at the few TeV scale, see Ref.~\cite{LQreview} for a general review on TeV LQs. 
The most economical case is the presence of a spin-1 state $U_1\sim ({\bf 3,1})_{2/3}$, however $R_{D^{(*)}}^{\tau\ell}$ requires a low mass scale, $m_*\sim 2\ {\rm TeV}$, being in tension with some observables, as the $Z$-couplings of third generation fermions~\cite{EFT_Azatov,us}. Besides, one can expect a whole set of spin-1 resonances at that scale, that generically induce too large $\Delta F=2$ operators, and eventually also large deviations in electroweak (EW) precision observables.  
Another possibility is the presence of two scalar LQs: a singlet $S_1\sim ({\bf \bar 3,1})_{1/3}$ and a triplet $S_3\sim ({\bf \bar 3,3})_{1/3}$. 
Several phenomenological analyses have shown the structure of couplings required for these states, most of them assuming interactions with Left-handed fermions only~\cite{Isidori,NP_Crivellin,NP_Angelescu}, but there are also some references that have included interactions with the Right-handed ones~\cite{NP_Bauer,Crivellin,NP_Crivellin2,Marzocca_2,Cata1}. 
Besides, as discussed in Ref.~\cite{1412.1791} and computed in Ref.~\cite{DaRold:2018moy} for $S_3$, if the scalars are Nambu Goldstone bosons (NGB) of the new sector, that become massive at loop level after interacting with the SM, their masses can be decoupled from the new sector scale: $m_S\ll m_*$, alleviating the bounds compared with the $U_1$ solution.

There have been a few attempts to obtain the scalar LQs from ultraviolet complete theories~\cite{NP_Marzocca,NP_Becirevic2}. 
We will consider that there is a new strongly coupled sector that generates resonances at a scale of order few tens of TeV, with the LQs and the Higgs emerging as NGBs, after spontaneous symmetry breaking by the strong dynamics.
Instead of the fundamental description, we will consider the effective weakly coupled theory of resonances, showing a coset that generates only these states as NGBs. The SM gauge bosons will gauge a subgroup of the global symmetry of the new sector and the SM fermions will be assumed to interact linearly with it. We will study, assuming flavor anarchy of the new sector, if it is possible to explain the $B$-anomalies, simultaneously passing flavor and EW bounds.

Our paper is organized as follows: in sec.~\ref{sec-model} we will describe the effective theory of resonances, the one loop potential of the scalars and the low energy theory, necessary to compute the contributions to flavor physics. In sec.~\ref{sec-observables} we will show the predictions of the model for the set of observables that receive the largest contributions, compared with the present bounds, and in sec.~\ref{sec-numerical} we will show the numerical predictions. In sec.~\ref{sec-resonances} we will describe the spectrum of fermion and vector resonances, and we will conclude. We leave some technical details for the appendices.

\section{A model with composite LQs and Higgs}\label{sec-model}
We are interested in the formulation of a model able to deliver the Higgs and the LQs $S_1$ and $S_3$ as pseudo-NGB (pNGB) resonances, generated by the spontaneous breaking of the global symmetry of an strongly coupled field theory (SCFT). 
The SM fermions and gauge bosons are assumed to be elementary fields, external to the strong dynamics, and weakly coupled to it. This is similar to the popular MCHM~\cite{Agashe-2004}, but with a larger set of light scalars, to include the LQs at a smaller scale than the rest of the resonances of the SCFT.

We assume that the SCFT has an exact global symmetry G, spontaneously broken by the strong dynamics to a subgroup H. After properly embedding the SM gauge symmetry into H, the spontaneously broken generators transform exactly as the Higgs boson and the required LQs. Besides these composite NGBs, the SCFT produces massive resonances characterized by a scale $m_*$, that we take of order 10-30~TeV. The currents associated to the global symmetry can create massive spin-1 states, transforming with the adjoint representation of G. We assume that the SCFT also produces fermionic massive resonances, generically at the same scale $m_*$. These fermions are assumed to transform with linear irreducible representations of G, realizing the symmetry at linear level. However, their representations are not fixed, leaving freedom for model building. All the resonances are taken to interact with typical couplings $g_*$, that are large compared with the SM ones, but still perturbative: $g_{\rm SM}\ll g_*< 4\pi$. For simplicity we take the decay constants of the different NGBs of the same order: 
\begin{equation}\label{eq-f}
f= m_*/g_* \ .
\end{equation}

The elementary gauge fields weakly gauge a subgroup of the global symmetry of the SCFT, interacting with the corresponding currents in the usual way. These interactions induce mixing between the elementary gauge fields and the spin-1 resonances, leading to interactions with the NGBs.

The elementary fermions are assumed to interact linearly with operators of the SCFT at a high energy scale: ${\cal L}\supset \omega \bar \psi{\cal O}^{\rm SCFT}$. Assuming approximate scale invariance, the running of the coupling $\omega$ is driven by the anomalous dimension of ${\cal O}^{\rm SCFT}$, leading to hierarchical couplings for different anomalous dimensions at low energy~\cite{Contino:2004vy}. At the scale $\sim m_*$ the operators create fermionic resonances and the linear interactions induce mixing with the elementary fermions, realizing partial compositeness~\cite{Kaplan:1991dc}:
\begin{equation}\label{eq-pc}
{\cal L}_{\rm mix}\supset \lambda f \bar \psi \Psi\ ,
\end{equation}
$\lambda$ is determined by $\omega(m_*)$ after rescaling. 

The interactions with the elementary fields explicitly break the global symmetry of the SCFT, generating a potential for the Higgs and LQs at 1-loop level. For that reason we will refer to them as pNGBs. Their masses can be estimated of order $m_{\rm pNGB}^2\sim g^2/(4\pi)^2\times m_*^2$, with $g$ the couplings between both sectors that explicitly break the global symmetries. The fermions can produce a negative Higgs mass squared, breaking dynamically  the EW symmetry.

\subsection{Global symmetries of the composite sector}
We consider the following coset G/H of the SCFT:
\be\label{eq-coset}
[\so{10}\tim\so{5}]/[\so{6}\tim\su{2}_A\tim\su{2}_B\tim\so{4}]
\te
that can deliver exactly $S_1$, $S_3$ and $H$ as pNGBs. Below we describe it in detail. We will use upper case letters for representations of G, and lower case letters for those of H.

The adjoint representation of a group G decomposes in representations of a subgroup H as 
$\rm{\bf{Adj}}(G) \to \rm{\bf adj}(H) \oplus \bf{r}_\Pi$, with ${\bf r}_\Pi$ the representation of the NGBs, that can be reducible. Considering that $\so{4} \sim \su{2}_C\tim\su{2}_R$, where the subindices are used to distinguish the different SU(2) subgroups in Eq.~(\ref{eq-coset}), the $\so{5}/\so{4}$ factor delivers a set of fields that transform as a singlet of $\so{6}\tim\su{2}_A\tim\su{2}_B$ and as a bidoublet of $\so{4}$, and can be identified with the Higgs as in the MCHM.

Regarding the other factor: $\so{10}/\so{6}\tim\su{2}_A\tim\su{2}_B$, the adjoint representation of SO(10) decomposes as
\be\label{eq-45-dec}
{\bf 45} \sim ({\bf 15},{\bf 1},{\bf 1})\oplus ({\bf 1},{\bf 3},{\bf 1}) \oplus ({\bf 1},{\bf 1},{\bf 3}) \oplus ({\bf 6},{\bf 2},{\bf 2}) \ , 
\te
where the first 3 representations correspond to the adjoint of the unbroken subgroup, and the last one to the broken generators. 

Joining the two factors we then have that the coset transforms as
\be\label{eq-rpi}
{\bf r}_\Pi = ({\bf 6},{\bf 2},{\bf 2},{\bf 1},{\bf 1}) \oplus ({\bf 1},{\bf 1},{\bf 1},{\bf 2},{\bf 2}) \equiv {\bf r}_{S} + {\bf r}_{H} \ .
\te

Now we discuss the embedding of the SM gauge symmetry, G$_{\rm SM}$, inside the subgroup H. The factor $\so{6} \sim \su{4}$ contains a subgroup $\su{3}_c\tim \uu{1}_X$, where we have identified the first factor with the color of G$_{\rm SM}$.  
Using that a ${\bf 6}$ of SU(4) decomposes under $\su{3}_c\tim \uu{1}_X$ as 
\begin{equation}\label{eq-6-dec}
{\bf 6}\sim{\bf 3}_{-2} \oplus {\bf \bar{3}}_{2}\ , 
\end{equation}
and that for SU(2) doublets: ${\bf 2} \otimes {\bf 2} \sim {\bf 3} \oplus {\bf 1}$, the singlet and triplet LQs, as well as the Higgs, can be obtained by identifying $\su{2}_L\equiv \su{2}_{A+B+C}$ and $Y=X/6+T^{3R}$. Notice that the embedding of $\su{2}_L$ is different from the usual one in SO(10) grand unification, since in that case SU(2)$_R$ is a subgroup of SO(10).

Under the unbroken global symmetry of the SCFT, both LQs are indistinguishable, as they are contained into a single irreducible representation. Since under SU(2)$_L$ they split in a triplet and a singlet, the weak interactions of the SM distinguish them.

\subsection{Representations of fermions}
The elementary fermions interact with the pNGBs after mixing with the fermionic resonances of the SCFT, Eq.~(\ref{eq-pc}). In order to preserve the local symmetry G$_{\rm SM}$, these resonances must transform with representations of the global symmetry of the SCFT that contain the representations of the SM fermions. For simplicity we assume that each elementary fermion mixes with just one operator, except in the case of leptons, where we explicitly consider two mixings: ${\cal L}\supset \bar l(\omega_{l_a}{\cal O}^{l_a}+\omega_{l_b}{\cal O}^{l_b})+\bar e(\omega_{e_a}{\cal O}^{e_a}+\omega_{e_b}{\cal O}^{e_b})$. We choose the same set of representations for all the generations. 

There is another requirement that guide us in the choice of the representations of the fermions of the SCFT: we demand Yukawa interactions with the pNGBs to reproduce the standard Higgs interactions, as well as interactions with the LQs needed for the phenomenology of the $B$-mesons.

We will consider the following SO(10) representations, and their decompositions under SO(6)$\times$SU(2)$_A\times$SU(2)$_B$:
\begin{align}\label{eq-repsSO10}
{\bf 16} &\sim ({\bf 4},{\bf 2},{\bf 1}) \oplus ({\bf \bar{4}},{\bf 1},{\bf 2}) 
 \ , \nonumber
\\
{\bf \overline{144}} &\sim {\bf (\bar{4},2,1)\oplus(4,1,2)\oplus(4,3,2)\oplus(\bar{4},2,3)\oplus(\overline{20},2,1)\oplus(20,1,2) } \ ,
\end{align}
as well as the fundamental representation of SO(5) and its decomposition under SO(4):
\be\label{eq-repsSO5}
{\bf 5\sim (2,2)\oplus (1,1)} \ .
\te
To follow the color charges of the fermions, it is also useful to know the following branching rules of SO(6) to SU(3)$_c\times$U(1)$_X$:
\begin{align}\label{eq-repsSU4}
\bf{4} \sim \bf 3_{\rm1} + 1_{\rm -3} \ , \qquad
\bf{15} \sim \bf 8_0 + 3_{4} + \bar 3_{4} + 1_{0} \ , \qquad
 {\bf 20} \sim \bf{ 3_{\rm1} + \bar{3}_{\rm5} + \bar{6}_{\rm1} + 8_{\rm -3} } \ .
\end{align}


In table \ref{t-rep-fermions} we define the representations of the fermionic operators of the SCFT and the resonances created by them. Each row is associated with an elementary fermion and the corresponding resonance mixing with it, as indicated by the subindices. On the first column we show the representations of the resonances under the full global symmetry G, in the second column we show the component under H containing the degrees of freedom with the same quantum numbers as the elementary fermions, while in the third and fourth columns we show the $X$ and $T^{3R}$ charges of the components mixing with the elementary fermions.
\begin{table}[ht!]
\centering
\begin{tabular}{|c|c|c|c|}
\hline\rule{0mm}{5mm}
SO(10)$\times$SO(5) & SO(6)$\times$SU(2)$^4$ & $X$ &$T^{3R}$
\\
\hline\rule{0mm}{5mm}
${\bf R}_{q}=({\bf 16},{\bf 5})$ & ${\bf r}_{q}$=({\bf 4},{\bf 2},{\bf 1},{\bf 1},{\bf 1}) & 1 & 0
\\
\hline\rule{0mm}{5mm}
${\bf R}_{u,d}=({\bf 16},{\bf 5})$ & ${\bf r}_{u,d}$=({\bf 4,2,1,2,2}) & 1 & $\pm1/2$
\\
\hline\rule{0mm}{5mm}
${\bf R}_{l_a}=({\bf \overline{16}},{\bf 5})$ & ${\bf r}_{l_a}$=({\bf 4,1,2,1,1}) & -3 & 0
\\
\hline\rule{0mm}{5mm}
${\bf R}_{e_a}=({\bf \overline{16}},{\bf 5})$ & ${\bf r}_{e_a}$=({\bf 4,1,2,2,2}) & -3 & -1/2
\\
\hline\rule{0mm}{5mm}
${\bf R}_{l_b}=({\bf \overline{144}},{\bf 5})$ & ${\bf r}_{l_b}$=({\bf 4,3,2,1,1}) & -3 & 0
\\
\hline\rule{0mm}{5mm}
${\bf R}_{e_b}= ({\bf \overline{144}},{\bf 5})$ & ${\bf r}_{e_b}$=({\bf 4,3,2,2,2}) & -3 & -1/2
\\
\hline
\end{tabular}
\caption{In the different columns we show the embeddings of the states with the same quantum numbers as the SM fermions, the rows indicate which elementary fermion mixes with them. We also show the charges $X$ and $T^{3R}$ of those components.}
\label{t-rep-fermions}
\end{table}

By making use of Eqs. (\ref{eq-repsSO10}-\ref{eq-repsSU4}) and the algebra of SU(2) it is straightforward to show that these massive resonances contain components with the same quantum numbers as the SM fermions. For ${\bf r}_{u,d}$ and ${\bf r}_{e_a}$ one must select the singlet contained in ${\bf 2}_A\otimes {\bf 2}_C$ and ${\bf 2}_B\otimes {\bf 2}_C$, respectively, whereas for ${\bf r}_{l_b}$ and ${\bf r}_{e_b}$ one chooses the doublet in ${\bf 3}_A\otimes {\bf 2}_B$ and the singlet in ${\bf 3}_A\otimes {\bf 2}_B\otimes {\bf 2}_C$, respectively, with the subindex labeling the corresponding SU(2) factors. 

Let us comment on several aspects of the chosen embedding. First, with respect to the SO(10) factor, it is possible to embed all the SM fermions either in ${\bf 16}$ or ${\bf 144}$ and one of their conjugates. 
Second, for the representation $({\bf\overline{144},5})$, besides the component $({\bf 4,3,2,1,1})$, it is also possible to embed $l_L$ in the component $({\bf 4,1,2,1,1})$, as in the case of $({\bf\overline{16},5})$. We will assume that the mixing with the components shown on the second column of Table~\ref{t-rep-fermions} dominates over the others, and for simplicity we will consider only those interactions. Third, the symmetry SU(2)$_A\times$SU(2)$_B$ does not allow diquark interactions of type $qqS_{1,3}$ and $duS_1$, we will elaborate more on this topic in the end of this section. 

The unbroken global symmetry allows the following interactions between resonances:
\begin{equation}\label{eq-Lcp1}
{\cal L}_*\supset y_{E*}\bar L H E + y_{U*}\bar Q \tilde H U + y_{D*}\bar Q H D + y_{3*}\bar Q^c\epsilon\sigma^a S_3^a L+y_{1*}\bar Q^c\epsilon S_1 L+y_{U*}\bar U^cS_1 E+ {\rm h.c.} \ ,
\end{equation}
where $y_*\sim g_*$ denote the couplings with the pNGBs, and $\epsilon=i\sigma_2$. Here we have used capital letters for the components of the multiplets of resonances that have the same quantum numbers as the elementary fermions, for example: if $\Psi_q$ transforms as ${\bf R}_q$, thus $\Psi_q=Q+\dots$, where $Q$ is the component that mixes with the elementary fermion $q_L$, meaning that it transforms as $({\bf 3,2})_{1/6}$ under G$_{\rm SM}$. A similar notation is assumed for the other states.

The embedding of the fermions can lead to relations between the couplings of the LQs. Embedding $L$ into ${\bf \overline{16}}$ of SO(10) leads to: $y_{1*}=y_{3*}$, whereas embedding it into ${\bf \overline{144}}$ gives: $y_{1*}=-3y_{3*}$, as is shown explicitly in Ap.~\ref{ap-cg}. The first case leads to a cancellation of the contributions to $R_{D^{(*)}}$ to leading order, instead one can consider either both embeddings, mixing $l_L$ with two resonances and obtaining independent linear combinations of couplings with the scalar LQs, or just the second one.

\subsection{Flavor structure: anarchic partial compositeness}\label{sec-apc}
Let us discuss now the flavor structure of the theory. We consider an anarchic SCFT, meaning that there is no flavor structure and all the flavor transitions are of the same order. In this case the Yukawa couplings of Eq.~(\ref{eq-Lcp1}) are tensors in flavor space that can be parametrized as: $y_{*ij}=g_*\times c_{ij}$, with all the coefficients of the anarchic matrices being of the same size: $c_{ij}\sim{\cal O}(1)$.

The hierarchy of Yukawa couplings needed to explain the masses and mixings of the SM fermions is generated by the structure of the mixing between the elementary and composite fermions: $\lambda$ in Eq.~(\ref{eq-pc}). Although $\lambda$ is a matrix in flavor space, it can be diagonalized by rotations of the elementary and composite fields~\cite{Panico:2015jxa}. For our work, it is enough to assume that $\lambda$ is diagonal and hierarchical.

The elementary-composite mixing can be diagonalized by performing a rotation of them, leading to partially composite massless fermions, that can be identified with the SM ones~\cite{2-site-sundrum}. We define their degree of compositeness as:
\begin{equation}
\epsilon=\frac{\lambda}{g_*}\ ,
\end{equation}
with $\epsilon\sim 1$ for a large degree of compositeness and $\epsilon\ll 1$ for mostly elementary fermions.
After EWSB these fermions become massive as in the SM. The Yukawa couplings with the Higgs are modulated by the mixing: $y_\psi\sim \epsilon_{\psi_L}g_*\epsilon_{\psi_R}$.

The hierarchy of masses and mixings of the quark sector can be reproduced by taking:
\begin{align}\label{eq-pcq}
& \epsilon_{q1}\sim\lambda_C^3\epsilon_{q3} \ ,
\qquad
\epsilon_{q2}\sim\lambda_C^2\epsilon_{q3} \ ,
\nonumber\\&
\epsilon_{u1}\sim \frac{m_u}{v_{\rm SM}}\frac{1}{\lambda_C^3g_*\epsilon_{q3}} \ ,
\qquad
\epsilon_{u2}\sim \frac{m_c}{v_{\rm SM}}\frac{1}{\lambda_C^2g_*\epsilon_{q3}} \ ,
\qquad
\epsilon_{u3}\sim \frac{m_t}{v_{\rm SM}}\frac{1}{g_*\epsilon_{q3}} \ ,
\nonumber\\& 
\epsilon_{d1}\sim \frac{m_d}{v_{\rm SM}}\frac{1}{\lambda_C^3g_*\epsilon_{q3}} \ ,
\qquad
\epsilon_{d2}\sim \frac{m_s}{v_{\rm SM}}\frac{1}{\lambda_C^2g_*\epsilon_{q3}} \ ,
\qquad
\epsilon_{d3}\sim \frac{m_b}{v_{\rm SM}}\frac{1}{g_*\epsilon_{q3}} \ .
\end{align}

For the sector of leptons there are two mixings per elementary multiplet, labeled by subindices $a,b$ in table~\ref{t-rep-fermions}. We will assume that $\epsilon_{\psi_a}\simeq\epsilon_{\psi_b}$, although it is also possible to consider the situation with $\epsilon_{\psi_a}\ll\epsilon_{\psi_b}$, and in the following we will not write this subindex anymore. Reproducing the masses of the charged leptons requires 
\begin{align}\label{eq-pcl}
\epsilon_{e1}\sim \frac{m_e}{v_{\rm SM}}\frac{1}{g_*\epsilon_{l 1}} \ ,
\qquad
\epsilon_{e2}\sim \frac{m_\mu}{v_{\rm SM}}\frac{1}{g_*\epsilon_{l 2}} \ ,
\qquad
\epsilon_{e3}\sim \frac{m_\tau}{v_{\rm SM}}\frac{1}{g_*\epsilon_{l 3}} \ ,
\end{align}
The mixings depend on the realization of neutrino masses, thus the relation between $\epsilon_{l i}$ is model dependent. Guided by the $B$-anomalies we will consider hierarchical Left-handed mixing: $\epsilon_{l 1}\ll\epsilon_{l 2}\ll\epsilon_{l 3}$, as well as $\epsilon_{ei}\ll\epsilon_{l i}$. In sec.~\ref{sec-numerical} we will show the numerical values favoured by flavor observables.

After integration of the fermionic resonances, at zero momentum, one obtains an effective Lagrangian
\begin{equation}\label{eq-LS1}
{\cal L}'_{\rm eff}\supset x_3\bar q_L^c\epsilon\sigma^a S_3^a l_L+x_1\bar q_L^c\epsilon S_1 l_L+x_{u}\bar u_R^cS_1 e_R + {\rm h.c.}
\end{equation}
where we only show the terms involving the LQs, similar interactions with the Higgs are present. If $\nu_R$ is included, there are new interactions containing this state. 

Anarchic partial compositeness generates a hierarchy of flavor in the LQ couplings, that is related with the hierarchy of the fermion masses:
\begin{align}\label{eq-couplings-pc}
x_{3,i\alpha}\sim g_*\ \epsilon_{qi}\ c_{3,i\alpha}\ \epsilon_{l\alpha} \ ,
\qquad
x_{1,i\alpha}\sim g_*\ \epsilon_{qi}\ c_{1,i\alpha}\  \epsilon_{l\alpha} \ ,
\qquad
x_{u,i\alpha}\sim c_{u,i\alpha}\frac{m_{u_i} m_{\ell_\alpha}}{\epsilon_{q i} \epsilon_{l \a} v_{\rm SM}^2 g_*} \ ,
\end{align}
where we have written $y_{n*}=g_*c_n$, for $n=1,3,U$.

\subsection{Constraints}
We consider now some general constraints, as proton stability and EW precision tests.

Grand unified theories usually lead to proton decay by exchange of LQs that also interact with diquarks, demanding a huge mass scale for these states. In the present model the LQs have masses $\lesssim{\cal O}(30)\ {\rm TeV}$, depending on whether they are pNGB as $S_1$ and $S_3$, or spin-1 resonances. However it is straightforward to show that, given the embeddings chosen for the fermions, at tree level there are no interactions of type $qq{\rm LQ}$. In fact there is a discrete symmetry that forbids those interactions, and makes the proton stable, a parity under which the quark resonances and $S_{1,3}$ are odd, whereas the leptonic resonances are even, allowing $qlS$ and forbidding $qqS$. One possibility to build such a transformation is by considering a $2\pi$ rotation with SU(2)$_A$, under which objects with half-integer spin, as quarks and $S$, are even, whereas objects with integer spin, as leptons, are singlets, as shown in Eq.~(\ref{eq-rpi}) and table~\ref{t-rep-fermions}. By demanding the elementary quarks (leptons) to be odd (even), this symmetry is preserved by the fermionic mixings. 

The previous symmetry does not forbid $n-\bar n$ oscillations, that can be mediated by dimension nine operators containing six fermionic resonances that mix with the quarks of the first generation~\cite{1410.1100}. However Ref.~\cite{1412.1791} has shown that in anarchic partial compositeness, with resonances in the TeV scale and couplings $g_*\sim 4\pi$, the bounds on the WCs of these operators can be satisfied. 

Below we discuss briefly the corrections to $Zb\bar b$ coupling and flavor observables arising from the presence of heavy massive resonances, the effect of the lighter pNGB LQs is considered in detail in the next section.

The $Zb_L\bar b_L$ coupling has been measured in agreement with the SM at a the level of $\sim 0.25\%$. Corrections in composite Higgs models characterized by one scale and one coupling can be estimated as $\delta g_{b_L}/g_{b_L}\sim \epsilon_{q3}^2 v_{\rm SM}^2/f^2$. As we estimate in sec.~\ref{sec-ewsb}, $v_{\rm SM}^2/f^2\sim 0.05$ for our benchmark region of parameters, thus for $\epsilon_{q3}\sim 1$ the bound is saturated. Although it is possible to protect this coupling with symmetries, the fermion embedding that we have chosen does not protect it, thus for the largest values of $\epsilon_{q3}^2v_{\rm SM}^2/f^2$ considered in this work some extra tuning should be present, whereas for the smallest values the estimate is an order of magnitude below the bound.

As is well known, meson phenomenology, as mixing and decays, put very strong limits on partial compositeness with flavor anarchy, demanding $m_*\gtrsim 10-30\ {\rm TeV}$~\cite{Csaki,Redi}. We take this scale for the resonances, at the price of increasing the amount of tuning demanded by the EW scale. Besides the mesons, the corrections to the neutron dipole moment and $\mu\to e\gamma$ require $f\gtrsim {\cal O}(5)\ {\rm TeV}$ and ${\cal O}(20-40)\ {\rm TeV}$, respectively~\cite{Panico:2015jxa}. There are different proposals to satisfy or alleviate these bounds~\cite{Csaki,Redi-Weiler,Panico-Pomarol}, most of them require departures from anarchy. However some interesting solutions for the lepton sector involving U(1) and CP symmetries have been discussed, for example, in Ref.~\cite{1807.04279}, whereas Ref.~\cite{1708.08515} has considered vanishing Right-handed mixing for the first generation: $\epsilon_{u1,d1,e1}\simeq 0$ and tiny bilinear interactions.

\subsection{Potential}
In order to estimate the masses of the LQs and analyse EWSB, it is useful to study the effective theory that contains the elementary fields and the NGBs, obtained after integration of the heavy resonances, and compute the potential induced at one loop. Since the structure of this effective theory is determined by the symmetries assumed for the underlying theory, it is possible to make some generic estimates without knowing more details of its dynamics~\cite{Callan:1969sn}.

The NGB unitary matrix is one of the main ingredients of this description. It is given by the exponential of the NGB fields $\Pi^{\hat a}$, with $\hat a$ running over the broken generators:
\begin{equation}
U\equiv e^{i\Pi/f} \ , \Pi=\Pi^{\hat a}T^{\hat a} \ ,
\end{equation}
with $T$ generators of the global symmetry group. $U$ transforms non-linearly under the action of an element ${\cal G}\in {\rm G}$: $U\to {\cal G}U{\cal H}^\dagger$, with ${\cal H}\in{\rm H}$ a function of ${\cal G}$ and $\Pi$. This transformation rule of $U$ is used extensively to build the effective theory.

The kinetic term of the NGBs is built by using the Maurer-Cartan form, defined as: $iU^\dagger D_\mu U=e^a_\mu T^a+d^{\hat a}_\mu T^{\hat a}$, with $a$ running over the unbroken generators, and $D_\mu$ being the covariant derivative that contains the elementary gauge fields. The kinetic term is: $(f^2/4)d_\mu^{\hat a}d^{\mu\hat a}$.

In this section we will consider only the fermions of the third generation, that, having the largest degree of compositeness, give the dominant contribution to the potential. Given that in our set-up the degree of compositeness of $b_R$ and $\tau_R$ are much smaller than the compositeness of the other fermions, we will not include them in this section.

In order to build the effective theory it is useful to promote the elementary fields to complete multiplets of the global symmetry of the SCFT: G=SO(10)$\times$SO(5). We do that by adding spurious elementary fields, that after the calculation must be set to zero. According to Table~\ref{t-rep-fermions}, we embed $q$ and $u$ in $\psi_{q,u}\sim({\bf 16,5})$, whereas for $l$ we consider two embeddings: $\psi_{l_a}\sim({\bf\overline{16},5})$ and $\psi_{l_b}\sim({\bf\overline{144},5})$. Only one linear combination of the components of $\psi_{l_a}$ and $\psi_{l_b}$, transforming as $(1,2)_{-1/2}$ of G$_{\rm SM}$, is dynamical.

The non-linear transformation properties of $U$ allow to build G-invariants that superficially look only H-invariant. To build them one has to dress the fermionic embeddings with the NGB matrix: $U^\dagger\psi$, decompose it under irreducible representations of H: ${\bf r}_H$, and build H-invariants by multiplication of dressed fields. At quadratic level in the elementary fermions {{and in momentum space}}:
%
\begin{align}\label{eq-LeffV}
&{\cal L}_{\rm eff}\supset \sum_f Z_f\bar\psi_f\pslash\psi_f+
\sum_{f,f'}\sum_{{\bf r}_H}\left[\Pi_{ff'}^{{\bf r}_H}(p)(\bar\psi_f U)P_{{\bf r}_H}(U^\dagger\psi_{f'})+\Pi_{ff'^c}^{{\bf r}_H}(p)(\bar\psi_f U)P_{{\bf r}_H}(U^\dagger\psi_{f'})^c\right]+{\rm h.c.} \nonumber 
\\
&f,f'=q,u,l_a,l_b \ ,
\end{align}
where $P_{{\bf r}_H}$ is a projector used for the decomposition on irreducible representations of H and the construction of H-invariants. The first term contains a kinetic contribution from the elementary sector. $\Pi_{ff'}^{{\bf r}_H}(p)$ is a singlet, it depends on momentum and codifies the information of the resonances that were integrated, its specific form depends on the realization of the SCFT or the theory of resonances, as for example discrete composite models or extra dimensions.

Armed with this effective Lagrangian, it is straightforward to compute the contribution of the elementary fermions to the Coleman-Weinberg potential at one loop:
\begin{equation}\label{eq-potentialCW}
V(\Pi)=-\frac{1}{2}\int\frac{d^4p}{(2\pi)^4} \log\det\ {\cal K}(\Pi) \ ,
\end{equation}
where ${\cal K}(\Pi)$ is the NGB-dependent matrix obtained by writing: ${\cal L}_{\rm eff}=\bar F{\cal K}(\Pi)F$, with $F={f,f^c}$, and $f=q,u,l$. 

\subsubsection{Masses of the LQs}
Expanding Eq.~(\ref{eq-potentialCW}) in powers of $\Pi$ one can obtain the masses of the LQs as momentum integrals of combinations of the correlators $\Pi_{ff'}^{{\bf r}_H}(p)$. More details are shown in Ap.~\ref{ap-potential}. We get: 
\begin{equation}\label{eq-MLQ1}
M_1^2=\tilde M^2+\Delta M_1^2 \ , 
\qquad
M_3^2=\tilde M^2+\Delta M_3^2 \ , 
\end{equation}
with $\tilde M^2$ and $\Delta M_{1,3}^2$ defined in Eq.~(\ref{eq-MLQAp}) of Ap.~\ref{ap-potential}.

The splitting between the LQs is driven by $\Psi_{l_b}$. For positive values of $M_1^2$ and $M_3^2$ there is no color breaking and Eq.~(\ref{eq-MLQ1}) gives the LQ masses to ${\cal O}(v^0)$. By noticing that $\C_{ff}^{{\bf r}_H}\sim{\cal O}(\epsilon_f^2)$, the order of these masses can be estimated as:
\begin{equation}
  M_{1,3}^2\sim \frac{g_*^2}{16\pi^2}m_*^2\epsilon_f^2 = (3\ {\rm TeV})^2\left(\frac{m_*}{20\ \rm TeV}\ \frac{g_*}{4}\ \frac{\epsilon_f}{1/2}\right)^2 \ .
  \label{mlq2}
\end{equation}

\subsubsection{Breaking of the EW symmetry}\label{sec-ewsb}
We start this analysis with some simple considerations about the gauge contributions to the Higgs potential. Since the Higgs arises as a NGB from the spontaneous breaking of the SO(5) factor, its interactions with the EW gauge bosons are as in the MCHM, the SO(10) factor does not play any role at one loop. The gauge contributions to the potential can be found, for example, in Refs.~\cite{Agashe-2004,Carena-2014}. The matching with the SM Higgs vacuum expectation value (vev) is given by
\begin{equation}
v_{\rm SM}^2=f^2 s_v^2 \ , \qquad s_v\equiv \sin(v/f)
\end{equation}
with $v=\langle h\rangle$ and $v_{\rm SM}\simeq 246$~GeV. For $m_*\simeq 20\ {\rm TeV}$ and $g_*\simeq4$ one gets: $s_v\simeq 0.05$, requiring a larger tuning than in the case of $f\sim 0.5-1\ {\rm TeV}$. 

The fermionic contribution to the potential can induce EWSB.
To study this breaking it is useful to evaluate Eq.~(\ref{eq-LeffV}) in the Higgs vev:
\begin{align}\label{eq-LeffVvev}
{\cal L}_{\rm eff}\supset 
\sum_{f=u,d,\nu,\ell} \bar f_L\pslash[Z_f+\hat\Pi_f(p)]f_L+
\bar u_R\pslash[Z_u+\hat\Pi_u(p)]u_R+ 
\bar u_L\hat M_u(p)u_R+ {\rm h.c.}\ .
\end{align}
The correlators $\Pi_f(p)$ and $M_f(p)$ can be obtained by matching ${\cal L}_{\rm eff}$ in the general background with the one with EWSB:
\begin{align}\label{eq-correlators}
& \hat\Pi_{f_L}=\sum_{{\bf r}_H} i_f^{{\bf r}_H}\ \Pi_{ff}^{{\bf r}_H}\ , 
\qquad f=u,d,\ ,
\nonumber \\
&
\hat\Pi_{u_R}=\sum_{{\bf r}_H} i_u^{{\bf r}_H}\ \Pi_{uu}^{{\bf r}_H} \ , 
\quad 
\hat M_u=\sum_{{\bf r}_H} j_u^{{\bf r}_H}\ \Pi_{qu}^{{\bf r}_H} \ ,
\nonumber \\
&
\hat\Pi_{f_L}=\sum_{{\bf r}_H} \left(i_{fa}^{{\bf r}_H}\ \Pi_{l_al_a}^{{\bf r}_H}+i_{fb}^{{\bf r}_H}\ \Pi_{l_bl_b}^{{\bf r}_H}\right) \ ,\qquad f=\nu,\ell \ .
\end{align}
The functions $i_f^{{\bf r}_H}(v)$ and $j_f^{{\bf r}_H}(v)$ contain the dependence with the Higgs vev, they are given in Table~\ref{t-inv-vev}, with $c_v\equiv\cos(v/f)$.
\begin{table}[ht!]
     \centering
     \begin{tabular}{|c|c|c|c|c|c|c|c|c|}
       \hline\rule{0mm}{5mm}
       H & $i_{u_L}$ & $i_{d_L}$ & $i_{u_R}$ & $i_{\nu_La}$ & $i_{\ell_La}$ &$ i_{\nu_Lb}$ & $i_{\ell_Lb}$ & $j_u$ \\[5pt]
       \hline\rule{0mm}{5mm}
                ${\bf r}_q $ & $c_v^2$ & $c_v^2$ & $\frac14 s_v^2$ &  0 & 0 &0&0 & $- \frac{i s_{2v}}{4} $ \\[5pt]
                \cline{1-9}\rule{0mm}{5mm}
                ${\bf r}_{u,d}$ & $s_v^2$ & $s_v^2$ & $\frac{7 + c_{2v}}{8}$ & 0 & 0 & 0 & 0 & $\frac{i s_{2v}}{4}$ \\[5pt]
                \cline{1-9}\rule{0mm}{5mm}
                ${\bf r}_{l_a}$ & 0 & 0 &  0 & $c_v^2$ & $c_v^2$ & 0 & 0 & 0  \\[5pt]
                \cline{1-9}\rule{0mm}{5mm}
                ${\bf r}_{e_a}$ & 0 &  0 & 0 & $s_v^2$ & $s_v^2$ & 0 & 0 & 0  \\[5pt]
                \cline{1-9}\rule{0mm}{5mm}
                ${\bf r}_{l_b}$ & 0 &  0 & 0 & 0 & 0 & $c_v^2$ & $c_v^2$ & 0  \\[5pt]
                \cline{1-9}\rule{0mm}{5mm}
                ${\bf r}_{e_b}$ & 0 &  0 & 0 & 0 & 0 & $s_v^2$ & $s_v^2$ & 0  \\[5pt]
                \hline
     \end{tabular}
     \caption{Fermionic invariants evaluated in the Higgs vev. The first column indicates the representation under SO(6)$\times$SU(2)$^4$, as defined in the second column of Table~\ref{t-rep-fermions}}
     \label{t-inv-vev}
\end{table}

Let us make a brief comment on the relation with the MCHM. The fermionic invariants are determined by the embedding of the elementary fermions in the larger symmetry of the SCFT. Given the identification of SU(2)$_L$, the SM doublets/singlets are embedded into singlets/doublets of SU(2)$_C$, as shown in Table~\ref{t-rep-fermions}. Thus, taking into account this subtlety and up to possible normalization factors, the $v$-dependence of the fermionic invariants can be obtained from the invariants of the MCHM with fermions in the fundamental representation.

Using the correlators it is straightforward to compute the Coleman-Weinberg potential that determines $v$ at one loop. Expanding in powers of $s_v$:
\begin{equation}\label{eq-potential-vev1}
V(v)\simeq-\alpha\ s_v^2+\beta\ s_v^4 \ ,
\end{equation}
where the quadratic and quartic coefficients can be written as integrals of the correlators, as shown in Ap.~\ref{ap-potential}, and are estimated as
\begin{equation}\label{eq-estimates-v4}
\alpha\sim \frac{N_c}{16\pi^2}m_*^4\epsilon_f^2 \ ,
\qquad
\beta\sim \frac{N_c}{16\pi^2}m_*^4\epsilon_f^4 \ .
\end{equation}
For $\epsilon_f$ one must take the dominating fermionic mixing, typically $f=q,u$ of the third generation, in the present case $f=l$ can also be large. 

The Higgs vev can be approximated by
\begin{equation}\label{eq-vSM-tuning}
s_v^2\simeq\frac{\alpha}{2\beta}\ ,
\end{equation}
requiring a tuning of order $1/s_v^2$ if we demand $v_{\rm SM}\ll f$. The different scaling of $\alpha$ and $\beta$ with $\epsilon_f$ has been considered in Ref.~\cite{1210.7114}, and is typical of fermionic embeddings in the fundamental and adjoint representations of SO(5). If the dominating $\epsilon_f$ is smaller than 1, it can lead to a problem of double tuning.

The Higgs mass can be estimated by using Eqs.~(\ref{eq-estimates-v4}) and (\ref{eq-vSM-tuning}) as
\begin{equation}\label{eq-estimate-mh}
m_h^2\simeq \frac{8}{f^2}\frac{\alpha(\beta-\alpha)}{\beta}
\sim \frac{N_c}{2\pi^2}g_*^4\epsilon_f^4 v_{\rm SM}^2 
\simeq \left[380\ {\rm GeV}\left(\frac{g_*}{4}\right)^2\left(\frac{\epsilon_f}{1/2}\right)^2\right]^2\ .
\end{equation}
Notice that, using Eq.~(\ref{eq-pcq}): $y_t\sim \epsilon_{q3}\epsilon_{u3}g_*$, thus taking similar mixings for both chiralities of the top quark and $g_*=4$, one obtains: $\epsilon_{q3,u3}\simeq 0.5$. In this case Eq.~(\ref{eq-estimate-mh}) denotes some tension with $m_h\sim 125\ {\rm GeV}$. Notice, in Eq.~(\ref{eq-ab-integrals}), that there are some ${\cal O}(1)$ factors inherited from Clebsch-Gordan coefficients, whose contributions to $\alpha$ are a factor 2-8 larger than in $\beta$, that can alleviate this tension. Calculations in explicit models of resonances, as for example in extra-dimensions~\cite{Agashe-2004,0612048} and discrete composite Higgs models~\cite{DeCurtis,Carena-2014}, have shown that this mass can be correlated with the presence of light fermionic resonances, usually $\lesssim 1\ {\rm TeV}$, also called custodians.

In the framework where one coupling $g_*$ and one scale $m_*$ characterize all the first level of resonances, fixing $m_*$ and $g_*$ and using Eq.~(\ref{eq-f}) one can obtain a lower bound for the tuning associated to (\ref{eq-vSM-tuning}). For $m_*\gtrsim 10-30\ {\rm TeV}$, as demanded by contributions of gluon resonances to $\epsilon_K$~\cite{Csaki}, taking $g_*\simeq 4$ leads to a tuning at least of order $(1-0.1)\%$.

If the potential is dominated by the fermionic contributions, when the fermionic resonances are lighter than the spin-1 resonances, some amount of tuning associated to Eq.~(\ref{eq-estimate-mh}) can be alleviated, since in this case the Higgs potential can be regulated by a lighter fermionic state~\cite{1210.7114}. Trading $m_*\to m_\psi=g_\psi f$, amounts to changing $g_*$ by $g_\psi$ in~(\ref{eq-estimate-mh}), that for $g_\psi=k_\psi g_*$, leads to a suppression factor $k_\psi^2$. For $k_\psi \simeq 0.5$ one can expect a Higgs mass of ${\cal O}(100\ {\rm GeV})$. A similar argument can be applied to the masses of the scalar LQs in Eq.~(\ref{mlq2}), in this case the masses are also rescaled by $k_\psi^2$, such that for $k_\psi \simeq 0.5$ one can expect masses of ${\cal O}(1\ {\rm TeV})$. On the other hand, the tuning from Eq.~(\ref{eq-vSM-tuning}) depends on the relative size of $\alpha$ and $\beta$, thus it is not expected to decrease with $k_\psi$, instead in the present model for $\epsilon_f<1$ one obtains a problem of double tuning. It is known that in this case the double tuning helps reducing the Higgs mass~\cite{1210.7114}.

\subsection{Low energy effective theory}
We consider now the effective theory at scales lower than the masses of $S_1$ and $S_3$.
Given the large value of $m_*\gg M_{1,3}$, we do not consider in our analysis the effect of the heavy spin-1 resonances on the low energy observables.
At low energies, integrating-out the LQs and Fierzing, leads to:
\begin{equation}\label{eq-LeffO}
{\cal L}_{\rm eff}^{\Lambda}\supset \sum \frac{C^i}{\Lambda^2} {\cal O}^i  + {\rm h.c.} \ ,
\end{equation}
with ${\cal O}^i$ given by:
\begin{align}\label{eq-Ops}
&{\cal O}^{T}_{\beta\alpha ij}=(\bar q_L^i\gamma_\mu\sigma^a q_L^j)(\bar l_L^\alpha\gamma^\mu\sigma^a l_L^\beta) \ , \qquad
{\cal O}^{1}_{\beta\alpha ij}=(\bar u_R^i q_L^j)\epsilon(\bar e_R^\beta l_L^\alpha) \ ,
\nonumber\\
&
{\cal O}^{S}_{\beta\alpha ij}=(\bar q_L^i\gamma_\mu q_L^j)(\bar l_L^\alpha\gamma^\mu l_L^\beta) \ ,
\end{align}
and 
\begin{align}\label{eq-matching-ceff}
&
\frac{C^{T}_{\beta\alpha ij}}{\Lambda^2} = -\frac{x_{1,i\beta}x^*_{1,j\alpha}}{4M_1^2}+\frac{x_{3,i\beta}x^*_{3,j\alpha}}{4M_3^2} \ ,
\qquad
\frac{C^{1}_{\beta\alpha ij}}{\Lambda^2} = \frac{x_{u,i\beta}x^*_{u,j\alpha}}{4M_1^2} \ ,
\nonumber \\
&
\frac{C^{S}_{\beta\alpha ij}}{\Lambda^2} = \frac{x_{1,i\beta}x^*_{1,j\alpha}}{4M_1^2}+3\frac{x_{3,i\beta}x^*_{3,j\alpha}}{4M_3^2} \ ,
\end{align}
where $i,j$ and $\alpha,\beta$ stand for generation indices of quarks and leptons, respectively.~\footnote{Notice that $C^{S,T}$ are normalized different from Ref.~\cite{Isidori}.}

Below the EW scale we rotate the fermions to the mass basis, replacing:
$\psi_X^i\to V^\psi_{X,ij}\psi_X^j$ for $\psi=u,d,\ell,\nu$ and $X=L,R$. For simplicity we choose the basis where $V^d_L=V^\ell_L=I$, the identity in three dimensions. It is straightforward to write the WCs of Eq.~(\ref{eq-matching-ceff}) in the new basis.

Other related operators that we are interested in, now written in the mass basis, are: 
\begin{align}
\mathcal{O}^{9 (10)}_{\ell_h \ell_i d_j d_k} &= \frac{\a}{4\pi} \left[\bar{d}_j \gamma^\mu P_L d_k\right]\left[\bar{\ell}_h \gamma_\mu \left(\gamma_5\right) \ell_i \right] \ , 
\label{eq-O910}\\ 
\mathcal{O}^7_{d_j d_k} &= \frac{e}{16\pi^2} m_k \left[\bar{d}_j \sigma^{\mu \nu} P_R d_k \right] F_{\mu\nu} \ , 
\label{eq-O7}\\
\mathcal{O}^{\rm VL (AL)}_{ \ell_h \nu_i u_j u_k} &= \left[\bar{u}_j \gamma^\mu (\gamma_5) d_k \right]\left[\bar{\ell}_h \gamma_\mu P_L \nu_i \right] \ , 
\label{eq-OVLAL}\\ 
\mathcal{O}^{L(R)}_{\ell_h \ell_i} &= \frac{e}{16\pi^2} \left[\bar{\ell}_h \sigma^{\mu \nu} P_{L(R)} \ell_i \right] F_{\mu \nu} \ , 
\label{eq-OLR}\\
\mathcal{O}^{SL(R)}_{ \ell_h \nu_i u_j u_k} &= \left[\bar{u}_j P_{L(R)}  d_k \right]\left[\bar{\ell}_h  P_L \nu_i \right] \ , 
\label{eq-OSLR}\\
{\cal O}^{dd}_{1,ij} &= (\bar d^i_L\gamma_\mu d^j_L)^2 \ .
\label{eq-Odd1}
\end{align}

In what follows we will denote by a capital letter $C$, with indices denoting the effective operator name, the NP contributions to the Wilson Coefficients. Otherwise, the SM contribution to these WC will be explicitly stated. 

For our estimations it is enough knowing that the masses of the LQs are of the same order, as discussed in the previous section, thus for simplicity we will take $M_1=M_3=M$. We find it useful to define:
\begin{equation}\label{eq-def-xi}
\d1\equiv\frac{g_*^2 v_{\rm SM}^2}{4M^2} \ .
\end{equation}
Looking at the definition of $\d1$ above, and taking into account the estimate for the LQ masses in Eq.~(\ref{eq-def-xi}), we expect $\d1$ to be approximately in the range $[0.02,0.3]$.

\section{Observables}\label{sec-observables}
In this section we analyse the impact of the new physics on low energy observables. We start with the so called $B$-anomalies: $R_{D^{(*)}}$ and $R_{K^{(*)}}$, and after them we consider constraints from other observables. We write the contributions in terms of the LQ couplings and then, making extensive use of partial compositeness, we show their dependence on the mixings, $\delta$ and $M$, as well as on the combinations of $c$ parameters defined in Eq.~(\ref{eq-couplings-pc}), that are taken of ${\cal O}(1)$. In the next section we will use these results for a combined numerical analysis of all the observables.

In this section we mostly follow the calculations of Refs.~\cite{Isidori} and~\cite{Crivellin}.

In what follows, expressions for $R_D^{(*)}$ and $R_{b\to c}^{\mu/e}$ actually refer to the ratio of its value with respect to the SM value, thus being equal to 1 in absence of NP contributions. 

\subsection{$R_{D^{(*)}}$}
Being a $b \to c \tau \nu$ process, this observable involves the operator $\O^{VL}-\O^{AL}$, that is generated at tree level by the LQ states. Following Ref.~\cite{Isidori}, we obtain:
\al
R_{D^{(*)}}^{\tau \ell} &\simeq 1+2  C^T_{3333} + 2 \frac{V_{tb}^*}{V_{ts}^*} C^T_{3323}  \nonumber 
\\  &\simeq 1 + \frac{v_{\rm SM}^2}{2} \left( \frac{|\xs{33}|^2}{M_1^2} - \frac{|\xt{33}|^2}{M_3^2} \right) + \frac{ V_{cs} v_{\rm SM}^2}{ 2 V_{cb}} \left( \frac{\xs{23} \xs{33}^* }{M_1^2} -  \frac{\xt{23} \xt{33}^*}{M_3^2}\right) \nn \\
&\sim 1 +2 \d1  \left\lbrace \epsilon_{q3}^2 \epsilon_{l3}^2 \left(|c_{1,33}|^2-|c_{3,33}|^2\right) + \frac{V_{cs}}{V_{cb}} \epsilon_{l3}^2 \epsilon_{q_2} \epsilon_{q3} (c_{1,23} c_{1,33}^* - c_{3,23}c_{3,33}^*) \label{rd:eq} \right\rbrace  \ . 
\fal
For the last estimate, that is valid up to coefficients of ${\cal O}(1)$, we have used partial compositeness. The coefficients $c_{i,jk}$ are of ${\cal O}(1)$, as discussed in sec.~\ref{sec-apc} they are assumed to be anarchic. Notice that all the corrections are of the same order given our flavor scheme. As reference value, we use $R_{D^{(*)},\rm exp}^{\tau \ell} = 1.14 \pm 0.057$. This is done by using HFLAV 2019 average \cite{hflavURL} to calculate the ratio of experimental to SM value, averaging between $R_D$ and $R_{D^*}$, due to the contribution being a symmetric one.

\subsection{$R_{K^{(*)}}$} 
This process requires a transition $b \to s \mu \mu$, involving the operator ${\cal O}^9 - {\cal O}^{10}$. These operators can be written in terms of $\mathcal{O}^T$ and $\mathcal{O}^S$, that in turn expressed as a function of the LQ couplings as:~\cite{Isidori}
\al
C^9_{2223} &= - C^{10}_{2223} =\frac{-\pi}{\a_{\rm em} V_{tb} V_{ts}^*} \left( C^T_{2223} + C^S_{2223} \right)  \nonumber \\ &= \frac{4 \pi}{\a_{\rm em} V_{tb} V_{ts}^*} \frac{v_{\rm SM}^2}{4 M_3^2} \xt{22} \xt{32}^* \nn \\
&\sim \frac{4 \pi}{\alpha_{\rm em} V_{tb} V_{ts}^*} \ \d1\ \epsilon_{l2}^2 \epsilon_{q_2} \epsilon_{q3}\	 c_{3,22} c_{3,32}^* \ .
\label{rk:eq}
\fal
The estimate of the third line is a consequence of the assumed flavor structure, and we show the dependence on the ${\cal O}(1)$ coefficients $c_{i,jk}$.

The fitted value consistent with experiment is~\cite{dC9}:
$ C^{9,\rm exp}_{2223} = -0.61 \pm 0.12$.

\subsection{$R_{b\to c}^{\mu/e}$}
This observable is related to $R_{D^{(*)}}$, and is also generated at tree level by the LQs. The main contribution is~\cite{Isidori}:
\al
R_{b\to c}^{\mu/e} - 1 &= 2 C^T_{2233} - 2 \frac{V_{tb}^*}{V_{ts}^*} C_{2223}^T \nonumber \\ &= \frac{v_{\rm SM}^2}{2} \left( \frac{|\xs{32}|^2}{M_1^2} - \frac{|\xt{32}|^2}{M_3^2} \right) + \frac{V_{cs} v_{\rm SM}^2 }{2 V_{cb}} \left(  \frac{\xs{22} \xs{32}^* }{M_1^2} - \frac{\xt{22} \xt{32}^*}{M_3^2}  \right) \nn \\
&\sim  2 \d1  \left\lbrace \epsilon_{q3}^2 \epsilon_{l2}^2 \left(|c_{1,32}|^2-|c_{3,32}|^2\right) + \frac{V_{cs}}{V_{cb}} \epsilon_{l2}^2 \epsilon_{q_2} \epsilon_{q3} (c_{1,22} c_{1,32}^* - c_{3,22}c_{3,32}^*) \right\rbrace   
\fal 
The experimental value is~\cite{PDG}: $R_{b\to c,\rm exp}^{\mu/e}-1=0.00 \pm 0.02$.

\subsection{$B_{K^{(*)}\nu \bar{\nu}}$} 
This observable also receives contributions at tree level in our model. The branching ratio of $B\to K^{(*)}\nu \bar{\nu}$, normalized to the SM, is~\cite{Isidori}:
\al
B_{K^{(*)}\nu \bar{\nu}} &= 1+ \frac23 \frac{\pi}{\alpha_{\rm em} V_{tb} V^*_{ts} C_{\nu}^{\rm SM}} \left( C^T_{3323} - C^S_{3323} + C^T_{2223} - C^S_{2223}  \right) \nn \\ &= 1 +   \frac23 \frac{\pi}{\alpha_{\rm em} V_{tb} V^*_{ts} C_{\nu}^{\rm SM}}  \frac{v_{\rm SM}^2}{2} \left( \frac{\xs{23} \xs{33}^*}{M_1^2} + \frac{\xt{23} \xt{33}^*}{M_3^2} +  \frac{\xs{22} \xs{32}^*}{M_1^2} + \frac{\xt{22} \xt{32}^*}{M_3^2} \right) \nn \\
&\sim 1 + \frac43 \frac{\pi}{\alpha_{\rm em} V_{tb} V^*_{ts} C_{\nu}^{\rm SM}} \d1 \epsilon_{l3}^2 \epsilon_{q_2} \epsilon_{q3} (c_{1,23} c_{1,33}^* + c_{3,23} c_{3,33}^*) + (l3 \to l2) \ ,
\fal
where $C_\nu^{\rm SM} = -6.4$.

The experimental constraint is $B_{K^{(*)}\nu \bar{\nu},\rm exp}< 2.6$~\cite{1702.03224}, {at 90\%CL}. 
 

By using the estimated values for $\d1$ above, we can check how relevant this bound is. For $\d1 \sim 0.02$, we have $B_{K^{(*)}} -1\sim 0.11 \left(2\epsilon\right)^4$, which does not greatly restrict the degree of compositeness of the third generation fermions. For larger values of $\d1$ this observable becomes more restrictive, but its importance still remains below that of other observables.

\subsection{$B \to K \tau \mu$ and $B_s \to \tau \mu$}
The scalar LQs induce $b \to s \tau \mu$ transitions that contribute to the decays $B \to K \tau \mu$ and $B_s \to \tau \mu$ with the operators of Eq.~(\ref{eq-O910}). 

We start with $B \to K \tau \mu$, in terms of their WCs~\cite{Crivellin,1504.07928}: 
\be
{\rm Br}\left[ B \to K \tau^{\pm} \mu^{\pm}  \right] = 10^{-9} \left\lbrace 9.6 \left( |C^9_{2323}|^2 + |C^9_{3223}|^2  \right)+10 \left( |C^{10}_{2323}|^2 + |C^{10}_{3223}|^2  \right) \right\rbrace \ .
\te 
With the contribution of $S_3$ to these WCs we have $C^9 = - C^{10}$, with:
\be 
C^9_{2323} = \frac{v_{\rm SM}^2 \pi}{V_{tb} V_{ts} \alpha M^2} \xt{32} \xt{23}^* \ .
\te 
Using the estimates of anarchic partial compositeness for the couplings we get: 
\be
{\rm Br}\left[ B \to K \tau^{\pm} \mu^{\pm}  \right] \sim 0.06 \, \d1^2 \epsilon_{q3}^4 \epsilon_{l2}^2 \epsilon_{l3}^2   \left(|c_{3,33}|^2 |c_{3,22}|^2  + |c_{3,32}|^2 |c_{3,23}|^2 \right) \ ,
\te
the experimental bound being $4.8 \times 10^{-5}${, at 90\%CL}. This observable is not expected to be too relevant under partial compositeness, as the combination above, for $\delta \sim 0.2$, $\epsilon_{q3},\epsilon_{l3} \sim 0.5$, $\epsilon_{l2}\sim 0.2$ gives $1.5\times 10^{-6}$, which is well below the bound. 

For $B_s \to \tau \mu$, using again $C^9 = -C^{10}$, one obtains~\cite{Crivellin}:
\be
{\rm Br}\left(B_s \to \tau \mu\right)= \frac{\alpha^2}{128 v^4 \pi^3} |V_{tb} V_{ts}|^2 f_{B_s}^2 \tau_{B_s} \left(m_\tau + m_\mu\right)^2 \eta(\frac{m_\tau}{m_{B_s}},\frac{m_\mu}{m_{B_s}}) |C^9_{3223}|^2 F(m_\tau,m_\mu) \ ,
\te
with $\eta(x,y) = \sqrt{1 - 2(x+y) + (x-y)^2}$, and 
\be
F = 1- \left(\frac{m_\tau - m_\mu}{m_{B_s}}\right)^2 +  \left(\frac{m_\tau - m_\mu}{m_\tau + m_\mu}\right)^2 \left( 1- \left(\frac{m_\tau + m_\mu}{m_{B_s}}\right)^2 \right) \ .
\te
Using $f_{B_s} = 0.225 \; {\rm GeV}, \tau_{B_s} = 1.47\times10^{-12}\,s, m_{B_s} = 5.36\; {\rm GeV}$ we get
\al
   {\rm Br}\left(B_s \to \tau \mu \right) &\simeq 5.3\times 10^{-3} V_{ts}^2 \left(\frac{\g {\rm TeV}}{M}\right)^4 \epsilon_{q3}^4 \epsilon_{l2}^2 \epsilon_{l3}^2 |c_{3,33}|^2 |c_{3,22}|^2 \nn \\
   &\sim 0.037 \, \d1^2 \, \epsilon_{q3}^4 \epsilon_{l2}^2 \epsilon_{l3}^2  \, |c_{3,33}|^2 |c_{3,22}|^2 \ .
\fal 

The experimental bound is: ${\rm Br}\left(B_s \to \tau \mu \right)_{\rm exp}\leq 4.2\times 10^{-5}${, at 95\%CL}. Same as above, we estimate the contribution to this observable to be $9.2\times10^{-7}$, which is also safely below the bound.

\subsection{$B_s \to \tau\tau$}
For this decay we have the branching ratio~\cite{Crivellin}
\be
\frac{{\rm Br}(B_s \to \tau \tau)}{{\rm Br}(B_s \to \tau \tau)_{\rm SM}} = \left\vert 1+ \frac{C^{10}_{3323}}{C^{10,\rm{SM}}_{3323}} \right\vert^2 \lesssim 8\times10^3 \ \quad\quad\quad {(95\%{\rm CL})}.
\te
The prediction for the ratio of coefficients is:
\al
\frac{C^{10}_{3323}}{C^{10,\rm SM}_{3323}} &\simeq - \frac{v_{\rm SM}^2 }{V_{ts} V_{tb} \alpha M_3^2}  \xt{33} \xt{23}^*
\sim 1700 \, \d1 \, \epsilon_{q3}^2 \epsilon_{l3}^2 \, c_{3,33}\, c_{3, 23}^* \ .
\fal

In this case, for the range of values of $\delta$, $\epsilon_{q3}$ and $\epsilon_{l3}$ of sec.~\ref{sec-model}, we estimate the ratio of the Wilson coefficients (WCs) to be of order $\sim \ \mathcal{O}(20)$, which gives a contribution to the branching ratio that in general is one order of magnitude below the bound, although in some cases it can reach the bound.

\subsection{$\tau \to \phi \mu$}
A contribution to this process is generated by $dd\ell\ell$ operators. The branching ratio can be expressed in terms of Left-handed couplings of $S_3$ as~\cite{1609.09078}:
\be
{\rm Br}\left(\tau \to \phi \mu \right) = \frac{f_\phi^2 m_\tau^3 \tau_\tau}{128 \pi} \frac{(\xt{22} \xt{23}^*)^2}{M_3^4} \left(1-\frac{m_\phi^2}{m_\tau^2}\right)\left( 1+2\frac{m_\phi^2}{m_\tau^2}\right) \ .
\te
We use $f_\phi = 0.225\;{\rm GeV}, m_\phi = 1.02\;{\rm GeV}$ and we get
\be
{\rm Br}\left(\tau \to \phi \mu \right) \sim 4 \times 10^{-6} \d1^2 \,  \epsilon_{q3}^4 \epsilon_{l2}^2 \epsilon_{l3}^2 \  c_{22}^2 \left(c_{23}^*\right)^2 \ .
\te

The experimental bound is $8.4\times10^{-8}${, at 90\%CL}. Typical values for the parameters give a contribution of $\sim \mathcal{O}(10^{-10})$.

\subsection{$B_c \to \tau \nu $} 
The branching ratio of $B_c \to \tau \nu $ can be expressed in terms of WCs as~\cite{1811.09603,1811.08899}:
\be
{\rm Br}\left(B_c \to \tau \nu \right) = 0.02 \left(\frac{f_{B_c}}{430 {\rm GeV}}\right)^2 |1+C^{VL}_{\tau \tau b c} + 4.3 (C^{SR}_{\tau \tau b c} - C^{SL}_{\tau \tau b c})|^2 \ .
\te
Both LQs contribute to $C^{VL}$, whereas only $S_1$ contributes to $C^{SL}$, as:
\begin{align}
C^{VL}_{\tau \tau cb}
&= \frac{-v_{\rm SM}^2}{4 V_{cb}} \sum_k\left( - \frac{V_{c k} \xs{k 3}^* \xs{33}}{M_1^2} + \frac{V_{c k} \xt{k 3}^* \xt{33}}{M_3^2}  \right) \nn
\\
&\simeq   \frac{-v_{\rm SM}^2}{4 V_{cb}} \left( - \frac{V_{cs} \xs{23}^* \xs{33} + V_{cb} |\xs{33}|^2}{M_1^2}  + \frac{V_{cs} \xt{23}^* \xt{33} + V_{cb} |\xt{33}|^2}{M_3^2}  \right) \ ,
\\
C^{SL}_{\tau \tau cb}& = \frac{-v_{\rm SM}^2}{4 V_{cb}} \frac{\xs{33} \xr{23}^*}{M_1^2} \ , 
\end{align}
while for Right-handed coefficients, without including $\nu_R$: $C^{SR} =0$. Besides, RGE running down, from the M$\sim$TeV, induces mixing between different WCs, such that the value of $C^{SL}$ gets corrected by an additional factor of 2.9, whereas the $C^{VL}$ coefficient has no correction~\cite{1706.00410}.

The estimates in our model are given by:
\begin{align}
& C^{VL}_{\tau\tau bc} \sim 1.5 \times 10^{-2} \left(\frac{\g {\rm TeV}}{M}\right)^2 \epsilon_{q3}^2 \epsilon_{l3}^2 (c_{3,33}c_{3,23}^* + |c_{3,33}|^2 - c_{1,33\tau} c_{1,23}^* -|c_{1,33}|^2) \ ,
\\
& C^{SL}_{\tau \tau b c} = \frac{-v_{\rm SM}^2}{4 V_{cb}} \frac{\xs{33} \xr{23}^*}{M^2} \sim - \frac{m_c m_\tau}{4 V_{ts} V_{cb} M^2} c_{1,33} \widetilde{c}_{1,23}^* \ .
\end{align} 

Combining both estimates we get:
\begin{align}
{\rm Br}\left(B_c \to \tau \nu \right) \simeq 0.02  \Big| 1 + 
&0.99 \, \d1 \,  \epsilon_{q3}^2 \epsilon_{l3}^2 \, (c_{3,33}c_{3,23}^* + |c_{3,33}|^2 - c_{1,33} c_{1,23}^* -|c_{1,33}|^2) \ , \nn
\\
& + 4.2 \times 10^{-3} \left(\frac{ {\rm TeV}}{M}\right)^2 c_{1,33} \widetilde{c}_{1,23}^* \Big|^2 \ .
\end{align}

This result has to be compared with an experimental bound ${\rm Br}\left(B_c \to \tau \nu \right)_{\rm exp} < 0.1${, at 90\%CL}. We do not expect this observable to play a significant role, as for the Left-handed contribution we estimate the branching ratio to be of $\sim 0.02$, while its Right-handed contribution gives also $\sim 0.02$ for $M\in [1,3]$ TeV. 

\subsection{$\Delta m_{B_s}$}
The contribution to this observable comes from the four-quark operator ${\cal O}^{dd}_1$ of Eq.~(\ref{eq-Odd1}), whose WC is generated at loop level by the scalar LQs, through a box diagram.
For the $B_s - \bar{B}_s$ system we have the following ratio:
\be
\frac{\Delta m_{B_s}}{\Delta m^{\rm SM}_{B_s}} = \Big\vert 1 + \frac{C^1_{sb}}{C^{1,\rm SM}_{sb}}\Big\vert
\te 
with the coefficients being~\cite{1008.1593}
\be
C_{sb}^{\rm1, SM} = 2.35 \frac{(V_{tb} V_{ts})^2}{ 8 \pi^2} \left( \frac{m_W}{v_{\rm SM}^2}\right)^2
\te
and~\cite{Crivellin}
\be
C^1_{sb} = \frac{1}{128 \pi^2 M^2} \left( \left(\xs{23}^*\right)^2 \xs{33}^2 + 5 \left(\xt{23}^*\right)^2\xt{33}^2 + 2 \xs{23}^*\xt{23}^* \xs{33}\xt{33} \right)
\te 
Among these three terms, when using anarchic partial compositeness, the one with the factor 5 will dominate the sum. We get 
\be
\frac{C^1_{sb}}{C^{1,\rm SM}_{sb}} \sim 300 \left(\frac{\rm TeV}{M}\right)^2 \, \d1^2 \epsilon_{q3}^4 \epsilon_{l3}^4  \left(c_{3,23}^* c_{3,33}\right)^2 \label{Dmb:eq}
\te

The most stringent bound is on the imaginary part of the WC~\cite{1302.0661}. We assume maximally violating phases of the LQ couplings, such that their effects on $\Delta m_{B_s}$ are restricted to be at most 20\% {(95\%CL)}. We expect this observable to play a role, as the value of the WC ratio for $\delta \sim 0.2$, $\epsilon_{q3}, \epsilon_{l3} \sim 0.5$ and $M \sim 2$ TeV, is $\sim 0.18$, which is close to the experimental limit. 

\subsection{Leptonic interactions of the $Z$}
We consider the flavor diagonal and flavor violating interactions of the $Z$ with charged leptons and neutrinos, that receive corrections at loop order, in particular we will be interested in the processes: $Z \to \tau_L \tau_L$, $Z \to \nu_L \nu_L$, $Z \to \tau \mu$, $Z \to \mu \mu$. We follow the results of Ref.~\cite{Crivellin}, see Ref.~\cite{Arnan} for the inclusion of subleading effects.

We consider the interaction terms at zero momentum transfer:
\al
\mathcal{L}_{\rm int}^{Z} = \frac{g}{c_W} \left[\left( \bar{\ell}_f \Gamma_{L,\ell_f \ell_i}(0) \gamma^\mu P_L \ell_i \right) + \lbrace L \to R \rbrace+\Gamma_{\nu_f \nu_i}(0)(\bar\nu_f\gamma^\mu P_L \nu_i)\right] Z_\mu  \ ,
\fal 
with $g$ the weak coupling and $c_W$ being the cosine of the Weinberg angle.
At one loop level the dominant corrections from the LQs are dominated by the contribution containing the top:
\begin{align}
&\Gamma_{L,\ell_f \ell_i} = \Gamma^{\rm SM}_{L,\ell_i} \delta_{fi} + \frac{N_c m_t^2}{32 \pi^2} \left[ \frac{ V_{3k} \xs{kf}^* V_{3l}^* \xs{li} }{M_1^2} \left( 1+\log\left(\frac{m_t^2}{M_1^2}\right) \right) 
+ \lbrace S_1 \to S_3, \xs{i\alpha} \to \xt{i\alpha} \rbrace 
\right] \ ,
\\
&\Gamma_{R,\ell_f \ell_i} = \Gamma^{\rm SM}_{R,\ell_i} \delta_{fi} - \frac{N_c m_t^2}{32 \pi^2}  \frac{ \xr{3f}^* \xr{3i} }{M_1^2} \left( 1+\log\left(\frac{m_t^2}{M_1^2} \right) \right) \ ,
\\
&\Gamma_{\nu_f \nu_i} = \Gamma^{\rm SM}_{\nu_i} \delta_{fi} + \frac{N_c m_t^2}{16\pi^2} \frac{V_{3k} \xt{kf}^* V_{3l}^* \xt{li}}{M_3^2} \left( 1+\log\left(\frac{m_t^2}{M_3^2} \right)\right) \ .
\end{align} 

Writing these flavor diagonal couplings in terms of the parameters of the model, we have:
\begin{align}
&\Gamma_{L,\tau \tau} - \Gamma_{L,\tau}^{\rm{SM}} \sim  -0.04 (|c_{1,33}|^2 + |c_{3,33}|^2)\, \d1 \, \epsilon_{q3}^2\, \epsilon_{l3}^2 \label{zttL:eq}\ ,
\\
&\Gamma_{R,\tau \tau} - \Gamma_{R,\tau}^{\rm{SM}} \sim -2 \times 10^{-8} \left(\frac{\rm{TeV}}{M}\right)^2 \frac{|\tilde{c}_{33}|^2}{\epsilon_{q3}^2 \epsilon_{l3}^2}\ ,
\\
&\Gamma_{L,\mu \mu} - \Gamma_{L,\mu}^{\rm{SM}} \sim  -0.04 (|c_{1,32}|^2 + |c_{3,32}|^2)\, \d1 \, \epsilon_{q3}^2\, \epsilon_{l2}^2 \ ,
\\
&\Gamma_{R,\mu \mu} - \Gamma_{R,\mu}^{\rm{SM}} \sim -6 \times10^{-11}    \left(\frac{\rm{TeV}}{M}\right)^2  \frac{|\tilde{c}_{32}|^2}{\epsilon_{q3}^2 \epsilon_{l2}^2}\ .
\end{align}
The Right-handed contributions to the couplings are heavily suppressed, the Left-handed contribution to the muon coupling is suppressed too, due to the small value for $\epsilon_{l2}$ compared with $\epsilon_{l3}$.
The SM predictions and the corresponding measurements can be found in Ref.~\cite{hep-ex/0509008}.

Regarding the $\delta \Gamma_\nu$ bound, there is a recent paper which gives an updated bound $N_\nu = 2.9963 \pm 0.0074$~\cite{Janot}. By using the relation
$N_\nu = 3 + 4 \, \delta \Gamma_\nu$, we get the bound
\be
\delta \Gamma_\nu = -0.000925 \pm 0.00185 \ .
\te
The expression in terms of the parameters of the model is:
\be
\delta \Gamma_\nu \sim  -0.09 \  |c_{3,33}|^2\, \d1 \, \epsilon_{q3}^2\, \epsilon_{l3}^2
\te 

For the $Z \to \tau \mu$ transition we have
\be 
{\rm Br}\left(Z\to \tau \mu \right) = \frac{K}{\Gamma_Z} \left( |\Gamma_{L,\tau \mu}|^2 + |\Gamma_{R,\tau \mu}|^2 \right) \ ,
\te
with $\Gamma_Z = 2.5\ {\rm GeV}$ the total Z width, and $K = 0.67\ {\rm GeV}$.
Replacing with the usual anarchic partial compositeness relations we have
\be
\Gamma_{L,\mu \tau} \sim -6.9 \times 10^{-4} \left(\frac{\g {\rm TeV}}{M}\right)^2 \epsilon_{q3}^2 \epsilon_{l2} \epsilon_{l3} \left(c_{1,33}c_{1,32}^*+c_{3,33}c_{3,32}^*\right)
\te
and
\be
\Gamma_{R,\mu \tau} \sim \frac{1.2 \times 10^{-9} \tilde{c}_{1,33}\tilde{c}_{1,32}^* \left(\frac{{\rm TeV}}{M}\right)^2}{ \epsilon_{q3}^2 \epsilon_{l2} \epsilon_{l3}} \ .
\te
Joining everything we obtain
\be
{\rm Br}\left(Z\to \tau\mu \right) \sim 2.8 \times 10^{-19} \, \frac{ \left(\tilde{c}_{1,33}\tilde{c}_{1,32}^*\right)^2 }{\left(\frac{M}{\rm TeV}\right)^4 \epsilon_{q3}^4 \epsilon_{l2}^2 \epsilon_{l3}^2} + 8.46 \times 10^{-6} \, \left(c_{1,33}c_{1,32}^*+c_{3,33}c_{3,32}^*\right)^2 \,  \d1^2  \epsilon_{q3}^4 \epsilon_{l2}^2 \epsilon_{l3}^2 \ ,
\te
to be compared against an experimental value of $1.2\times10^{-5}${, at 95\%CL}. Looking at its expression, we see that for values of order $\epsilon_{l3} \sim 0.5$, $\epsilon_{q3}\sim 0.5$, $\epsilon_{l2} \sim 0.1$, $M\sim \rm{TeV}$, the Right-handed contribution to this branching ratio is heavily suppressed with respect to the Left-handed one.

\subsection{{$\ell_i \to \ell_f \gamma$}}\label{sec-eiefgamma}
These flavor violating decays are produced by operators ${\cal O}^{L,R}_{\ell_h \ell_i}$ of Eq.~(\ref{eq-OLR}). 
Following Ref.~\cite{Crivellin}, the LQs give a contribution to the WCs of these operators at one loop level that can be written as:
\be
C^L_{\ell_f \ell_i} = - \frac{m_{\ell_f} \xs{3 f}^* \xs{3 i} + m_{\ell_i} \xr{3f}^* \xr{3 i}}{8 M_1^2} + \frac{m_t \xr{3f}^* V^*_{3k} \xs{ki}}{4 M_1^2}\left(7+4 \log\left(\frac{m_t^2}{M_1^2}\right)\right) + \frac{3 m_{\ell_f} \xt{3f}^*\xt{3i}}{8 M_3^2}
\te 
with $C^R = {C^L}^\dagger$, due to hermiticity. The branching ratio for the transition is written as 
\be 
{\rm Br}(\ell_i \to {\ell_f} \gamma ) = \frac{\alpha m_{\ell_i}^3 \tau_{\ell_i}}{256 \pi^4} \left( |C^L_{\ell_f \ell_i} |^2 + |C^R_{\ell_f \ell_i}|^2\right)
\te

We want to estimate the size of the transition $\tau \to \mu \gamma$, and $\mu \to e \gamma$
For the first one, supposing only Left-handed $S_3$ couplings dominate, we get
 \be
{\rm Br}(\tau \to \mu \gamma) \sim 1.4 \times 10^{-3} \ \d1^2 \,  \epsilon_{q3}^4 \epsilon_{l2}^2 \epsilon_{l3}^2 |c_{3,33}|^2 |c_{3,32}|^2 \ ,
\label{taumgL:eq} 
\te 
whereas if Right-handed couplings dominate, we get
\be
{\rm Br}(\tau \to \mu \gamma) \sim 1.7 \times 10^{-6} \left(\frac{{\rm TeV}}{M}\right)^4 \frac{\epsilon_{l2}^2}{\epsilon_{l3}^2} \left(\widetilde{c}_{33}^2 c_{1,32}^2 + \left(\frac{m_\mu}{m_\tau}\right)^2 \left(\frac{\epsilon_{l3}}{\epsilon_{l2}}\right)^4\widetilde{c}_{32}^2 c_{1,33}^2\right) \ .
\label{taumgR:eq}
\te
For this contribution to the branching ratio, we note first that it has an explicit dependence on $M$ that goes like $M^{-4}$, thus, the contribution grows for smaller values of the LQ masses. Also, we recognize two regimes that contribute to this quantity. For $\epsilon_{l2} \gtrsim \epsilon_{l3}$, the second term is suppressed by the ratio of muon to tau mass. For $\epsilon_{l3} \gtrsim \sqrt{\frac{m_\tau}{m_\mu}} \epsilon_{l2}$, the second term starts to dominate.

The experimental bound from Ref.~\cite{0908.2381} is: ${\rm Br}(\tau \to \mu \gamma)_{\rm exp} < 4.4 \times 10^{-8}${, at 90\%CL}.

{{{In the case of $\mu \to e \gamma$, we use the expressions above, changing the lepton flavors and $m_e\sim 511\ {\rm keV}$, $\tau_{\mu} \sim 2\ \mu s$. For the Left-handed contribution, we get an expression similar to Eq.(\ref{taumgL:eq}),
 \be
{\rm Br}(\mu \to e \gamma) \sim 7 \times 10^{-3} \ \d1^2 \,  \epsilon_{q3}^4 \epsilon_{l1}^2 \epsilon_{l2}^2 |c_{3,32}|^2 |c_{3,31}|^2 \ ,
\label{muegL:eq}
\te 
Whereas for Right-handed couplings the contribution is
\be
{\rm Br}(\mu \to e \gamma) \sim 8 \times 10^{-6} \, \left(\frac{{\rm TeV}}{M}\right)^4 \frac{\epsilon_{l1}^2}{\epsilon_{l2}^2} \left(\widetilde{c}_{32}^2 c_{1,31}^2 + \left(\frac{m_e}{m_\mu}\right)^2 \left(\frac{\epsilon_{l2}}{\epsilon_{l1}}\right)^4\widetilde{c}_{31}^2 c_{1,32}^2\right) \ .
\label{muegR:eq}
\te
The experimental bound is ${\rm Br}(\mu \to e \gamma) < 4.2 \times 10^{-13}$~\cite{MEG}, at 90\%CL. The Left-handed contribution, taking as an example similar degree of compositeness of both chiralities of the electron, $\epsilon_{l1}\sim\epsilon_{e1} \sim 7 \times 10^{-4}$, and other typical values for the parameters, is of order $3.6 \times 10^{-15}$. The Right-handed contribution, however, is a bit more compromised. Eq.(\ref{muegR:eq}) has a minimum for $\epsilon_{l1}/\epsilon_{l2}\sim 0.07$, leading to ${\rm Br}(\mu \to e \gamma)\sim 4\times 10^{-8}({\rm TeV}/M)^4$. In this setup, for $M=1\ {\rm TeV}$ a cancellation of order $10^{-5}$ is required, otherwise $M\gtrsim 20\ {\rm TeV}$. Another possibility is to decouple the electron mass from partial compositeness, assuming that its degree of compositeness is much smaller than the previous estimates and that its mass is generated by anarchic tiny bilinear interactions of the elementary fermions with the Higgs~\cite{1708.08515} (see also \cite{Matsedonskyi,Panico-Pomarol} for other related approaches). In the following we will assume this to be the case.}}

\subsection{{{$\ell_i \to 3 \ell_f$}}}
{{We consider here observables $\tau \to 3 \mu$ and $\mu \to 3 e$, which have loop level contributions, induced by the flavor violating $Z\mu\tau$ and $Ze\mu$ couplings, and four-lepton operators~\cite{Isidori,1705.00929}:
\al
{\rm Br}\left(\ell_i \to 3\ell_f \right) 
&= 2.5\times 10^{-4} \left(C^T_{if33} - C^S_{if33}\right)^2 \nn \\ &= 6.25 \times 10^{-5} \left( \frac{v_{\rm SM}^2 \xs{3i} \xs{3f}^*}{M_1^2} +   \frac{v_{\rm SM}^2 \xt{3i} \xt{3f}^*}{M_3^2}   \right)^2 \nn \\
&\sim 0.001 \,  \d1^2 \epsilon_{q3}^4 \epsilon_{lf}^2 \epsilon_{li}^2 \left(c_{1,3f}^* c_{1,3i} + c_{3,3f}^* c_{3,3i} \right)^2 \ .
\label{tau3m:eq}
\fal
For $\tau \to 3 \mu$ decay, we have $i\to3$, $f\to2$ in the expression above. The experimental bound is ${\rm Br}\left(\tau \to 3\mu\right)_{\rm exp}< 1.2 \times 10^{-8}$, at 90\% CL. This value, along with the $\tau \to \mu \gamma$ are expected to increase in sensitivity by an order of magnitude in Belle II~\cite{1808.10567}.}

{{For the $\mu \to 3 e$ decay, we set $i\to 2$, $f \to 1$. The experimental limit is, at 90\% CL is ${\rm Br}\left({\mu \to 3 e}\right)_{\rm exp} < 1.0 \times 10^{-12}$~ \cite{mu3e}.
The expected size of this observable now depends on the size of the mixing to first generation leptons, $\epsilon_{l1}$. For $\epsilon_{l1}\sim\epsilon_{e1}$, $\d1 \sim 0.02-0.2$, $\epsilon_{q3} \sim 0.5$ and $\epsilon_{l2} \sim 0.2$, the size of this branching ratio is at least two orders of magnitude below the experimental limit. For non-symmetric mixing $\epsilon_{l1}$ can be taken of order $0.003$ or $0.03$ if $\d1 \sim 0.2$ or $0.02$, respectively. In the case of negligible linear mixing this process does give interesting constraints.}

\subsection{LFU in $W$ couplings}
The LQs generate contributions to $W$ couplings at one loop that violate lepton universality. In the present model the relevant modifications are for the leptons of the third generation~\cite{Isidori,1705.00929}:
\al
\left\vert\frac{g^W_\tau}{g^W_\ell}\right\vert &= 1- 0.084 \, C^T_{3333}=  1-0.084 \left(\frac{v_{\rm SM}^2}{4 M_1^2} |\xs{33}|^2 - \frac{v_{\rm SM}^2}{4 M_3^2} |\xt{33}|^2\right) \nn \\
&\sim 1 - 0.084 \, \d1 \, \epsilon_{q3}^2 \epsilon_{l3}^2 \left(|c_{1,33}|^2-|c_{3,33}|^2\right)
\fal

The ratio $|g^W_\tau/g^W_\ell| $ is measured to be $1.0000 \pm 0.0014$~\cite{pich}{, at 95\%CL}.

\section{Numerical results for flavor physics}\label{sec-numerical}
We wish to test if the $B$-anomalies and the flavor constraints detailed above can be made compatible with an anarchic partial compositeness scenario. For this purpose we will explore {{if $R_{K^{(*)}}$ and $R_{D^{(*)}}$ can be fitted simultaneously to within $1\sigma$ of their experimental values, with the bounds being satisfied at the confidence levels specified in the previous section}}. 

The observables depend on different combinations of the parameters $c_{1,i \alpha}, c_{3,i\alpha}$ and $\tilde{c}_{1,i\alpha}$, we will refer to those combinations as $\Delta_O^{(i)}$, with $O$ the observable, and $i$ an index labelling the number of independent combinations of that $O$. For example, for $R_{D^{(*)}}$ we have the combinations:
\al
 \Delta^{(1)}_{R_{D^{(*)}}}\equiv |c_{1,33}|^2 - |c_{3,33}|^2 \ , \qquad
 \Delta^{(2)}_{R_{D^{(*)}}}\equiv c_{1,23}c_{1,33}^* - c_{3,23} c_{3,33}^* \ .
\fal
For each observable that has a different combination of the parameters $c_{1,i \alpha}, c_{3,i\alpha}$ or $\tilde{c}_{1,i\alpha}$, as we are working under the assumption of flavor anarchy, we will take all these coefficients as independent and of the same order. For particular values of these coefficients, the model can pass all flavor constraints and simultaneously explain the $B$-anomalies, however, we will explore whether this happens for generic ${\cal O}(1)$ coefficients. Whenever some $\Delta_O$ is required to deviate from ${\cal O}(1)$, the assumption of anarchic partial compositeness is in tension with that observable. Typically, the bounds from flavor observables are expected to favor $\Delta_O<{\cal O}(1)$, showing the need of some alignment or tuning, since in the limit of vanishing $\Delta_O$ the new physics contributions vanish. On the other hand, an explanation of the $B$-anomalies requires sizable $\Delta_{R_{D^{(*)}}}$ and $\Delta_{R_{K^{(*)}}}$, and for some regions of the parameter space they can be required to be: $\Delta_O>{\cal O}(1)$, deviating from the assumption of flavor anarchy. 

Besides $\Delta_O$, the observables depend on $\d1$, defined in Eq.~(\ref{eq-def-xi}), on the LQ mass $M$, as well as on the Left-handed mixings of the top and the leptons, since we have used Eqs.~(\ref{eq-pcq}) and (\ref{eq-pcl}) to fix the size of the other mixings. 

To estimate the amount of tuning one expects in the $\mathcal{O}(1)$ coefficients contributing to the $B$-anomalies, we proceed in the following way: for a given point of the parameter space, we compute which are the values of $\Delta_{\mathcal{O}}$ that cause the observables to fall within the $1\sigma$ experimental value {{and the corresponding CL intervals}}. In those cases where there is more than one $\Delta_O$, as in $R_{D^{(*)}}$ or $\tau \to \mu \gamma$, we consider either the largest contribution, if they are of different order, or consider them separately, if their ordering depends on the particular region of the parameter space. Then we select the points that can reproduce all the flavor bounds with $\Delta_O$ of order 1, allowing for a certain threshold. To do this, we perform a random scan over the free parameters, we take: $\epsilon_{q3},\epsilon_{l3}\in [0.5,1]$, $\epsilon_{l2} \in [0.08,0.25]$, $\delta \in [0.02,0.2]$, and $M \in [1,3]$ TeV. Scanning over $200k$ initial points, we select the ones that have $\rm{min}\left(\Delta_O\right) \geq 0.95$, obtaining $\sim 10k$ points. There are 4 observables that have the smallest $\Delta_O$, and can thus be identified as the most sensitive, these are: {{$\Delta m_{B_s}$ on 35\% of the points, followed by $g^W_\tau$ on 28\% of the points, followed by $\tau \to \mu \gamma _{\rm(R)}$ on 23\% of the points and by $\tau \to 3 \mu$ with the remaining 14\%}}. These are the observables that impose the most stringent bounds on the parameter space. Then, to estimate the amount of tuning required to explain the $B$-anomalies, we plot the required values for $\Delta_{R_{D^{(*)}}}$ and $\Delta_{R_{K^{(*)}}}$ on these points. We show our results in the distribution of Fig.~\ref{hist2d}, where we have truncated the upper limits of the graph to have a better focus on its densest region, as the tails of the distribution go to higher values but with a very small density. In this figure we see that explaining $R_{D^{(*)}}$ at $1\sigma$ level requires some tuning, since the peak of $\Delta_{R_{D^{(*)}}}$ is in the range {{$3-6$}}, whereas $R_{K^{(*)}}$ can be explained with {{$\Delta_{R_{K^{(*)}}}\sim 0.25-1$}}. This result shows that the former observable is in tension with the flavor constraints. Similarly, by allowing for higher tuning in the flavor constraints, that is: allowing their $\Delta_{O}\sim 0.05-0.3$, one can take $\Delta_{R_{D^{(*)}}}\simeq 1$. {{Besides we find that the distribution of $M$ is peaked around $1.8\ {\rm TeV}$.}}

\begin{figure}[h!]
  \centering
  \includegraphics[width=0.99\textwidth]{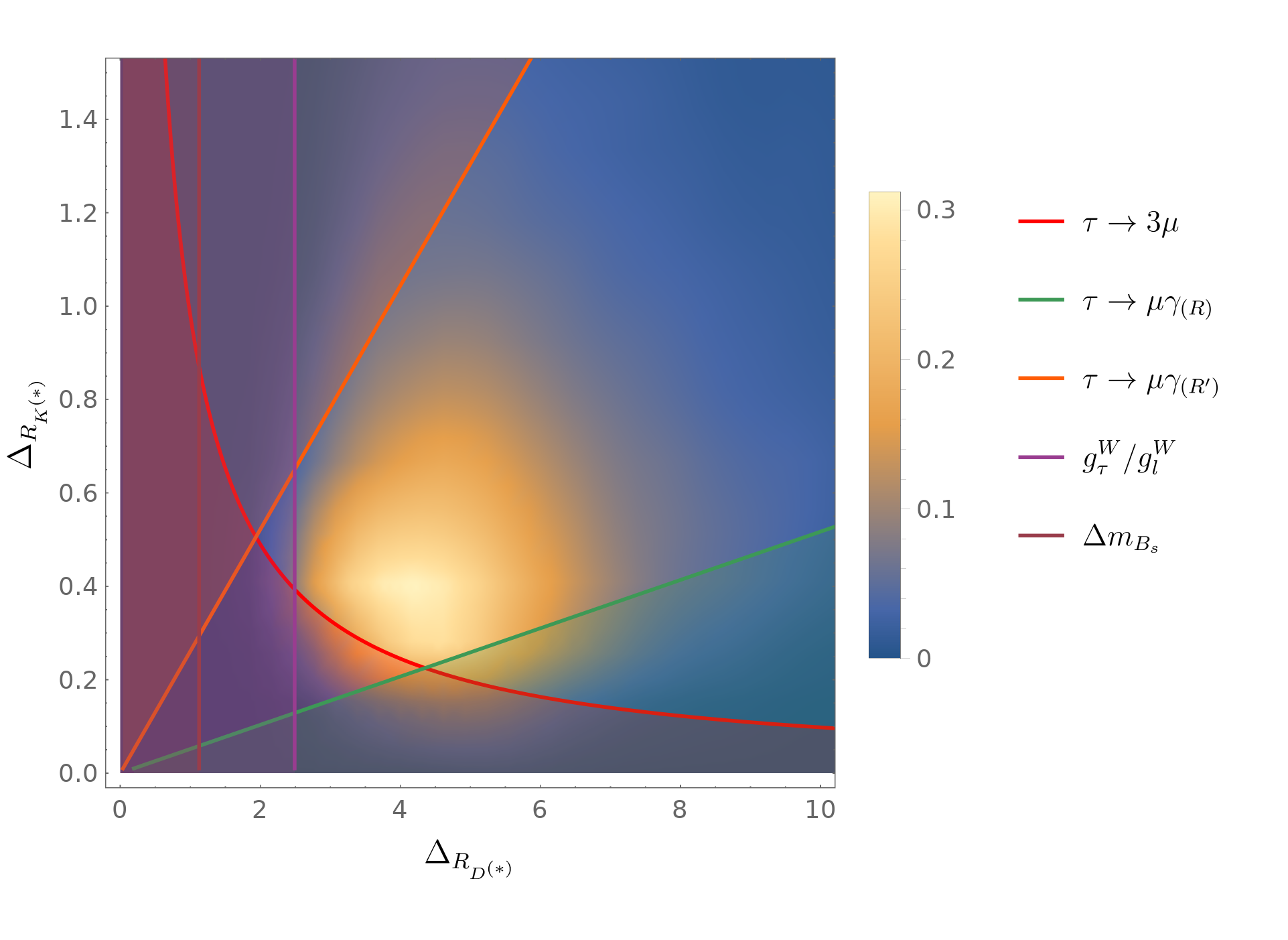}
  \caption{Distribution of required values for $\Delta_{R_{D^{(*)}}}$ and $\Delta_{R_{K^{(*)}}}$, for points passing all flavor observables with the other $\Delta_O \simeq 1$. {The colored curves show the estimates of the bounds coming from Eq.~(\ref{eq-rel1}) and below, for $M=1.5$ TeV. The regions excluded following those approximations have been shaded.}}
  \label{hist2d}
\end{figure}

We can explain the shape of the lower limit of this region by looking at what flavor constraint those points correspond to. {{Let us consider}} $\tau \to 3 \mu$, we have to check this observable's expression along with those of $R_{K^{(*)}}$ and $R_{D^{(*)}}$. We see that $\tau \to 3\mu$ depends on 4 of the 5 parameters in the random scans. Furthermore, we can multiply the expressions for $R_{K^{(*)}}$ and $R_{D^{(*)}}$, taking into account the lower $1\sigma$ limits for the observables. In this product, we then replace the combination of parameters $\delta^2 \epsilon_{q3}^4 \epsilon_{\l2}^2 \epsilon_{l3}^2$ {that saturates the bound} in $\tau \to 3 \mu$, getting:
{{
\be\label{eq-rel1}
\Delta_{R_K^{(*)}} \Delta_{R_D^{(*)}} \simeq \frac{0.49 \times 0.083 \, V_{tb} \, \alpha_{\rm em}  \Delta_{\tau 3 \mu}}{1.2 \times 10^{-5} \, 8 \pi \,} \simeq \, \Delta_{\tau 3 \mu} \ .
\te
Which partially explains the shape of the lower limit as seen in Fig.~\ref{hist2d} above. Making a similar analysis with $g_\tau^W$ we obtain: $\Delta_{R_{D^{(*)}}}\gtrsim 2.5\Delta_{g_\tau^W}$. The other observables depend also on $M$, from $\Delta m_{B_s}$ we obtain: $\Delta_{R_{D^{(*)}}}\gtrsim 1.6 M\Delta_{\Delta m_{B_s}}^{1/2}$, and from $\tau\to\mu\gamma$ we obtain a lower limit $\Delta_{R_{D^{(*)}}}\gtrsim 18 ({\rm TeV}/M)^4\Delta_{R_{K^{(*)}}}\Delta_{\tau\to\mu\gamma}^{(R')}$ and a lower limit $\Delta_{R_{D^{(*)}}}\lesssim 3.5 (M/{\rm TeV})^4\Delta_{R_{K^{(*)}}}/\Delta_{\tau\to\mu\gamma}^{(R)}$, where the superindices indicate the different combinations of coefficients present in the Right-handed contribution to this process. For $M\lesssim 1.3\ {\rm TeV}$ the bounds from $\tau\to\mu\gamma$ are not compatible with $\Delta\simeq 1$, whereas $g_\tau^W$ and $\Delta m_{B_s}$ give a lower bound $\Delta_{R_{D^{(*)}}}\gtrsim 2.5$.
}}

The previous results show that there is a minimum amount of tuning, since $\Delta_{R_{D^{(*)}}}$, expected to be ${\cal O}(1)$, must be of {{${\cal O}(2.5-7)$}} when the other $\Delta_{\cal O}$ are of ${\cal O}(1)$, requiring some alignment or tuning. We will consider a scenario referred as minimal tuning, in which $\Delta_{R_{K^{(*)}}},\Delta_{R_{D^{(*)}}}\leq 5$, whereas $\Delta_{\cal O}\geq 0.3$.
In the case of the $B$-anomalies, we plot a contour line for those points that require $\Delta_{R_{D^{(*)}}}=5$ and/or $\Delta_{R_{K^{(*)}}}=5$ in order to explain $R_{D^{(*)},\rm exp}$ or $R_{K^{(*)},\rm exp}$, and we show in green the region where any of those $\Delta_O$ are required to be larger than 5. 
For the other observables, this is done by plotting a contour line with $\Delta_O = 0.3$, while in red we show the region with $\Delta_O<0.3$. In the white region the observables can be reproduced with minimal flavor tuning.

\begin{figure}[h!]
  \centering
  \includegraphics[width=0.495\textwidth]{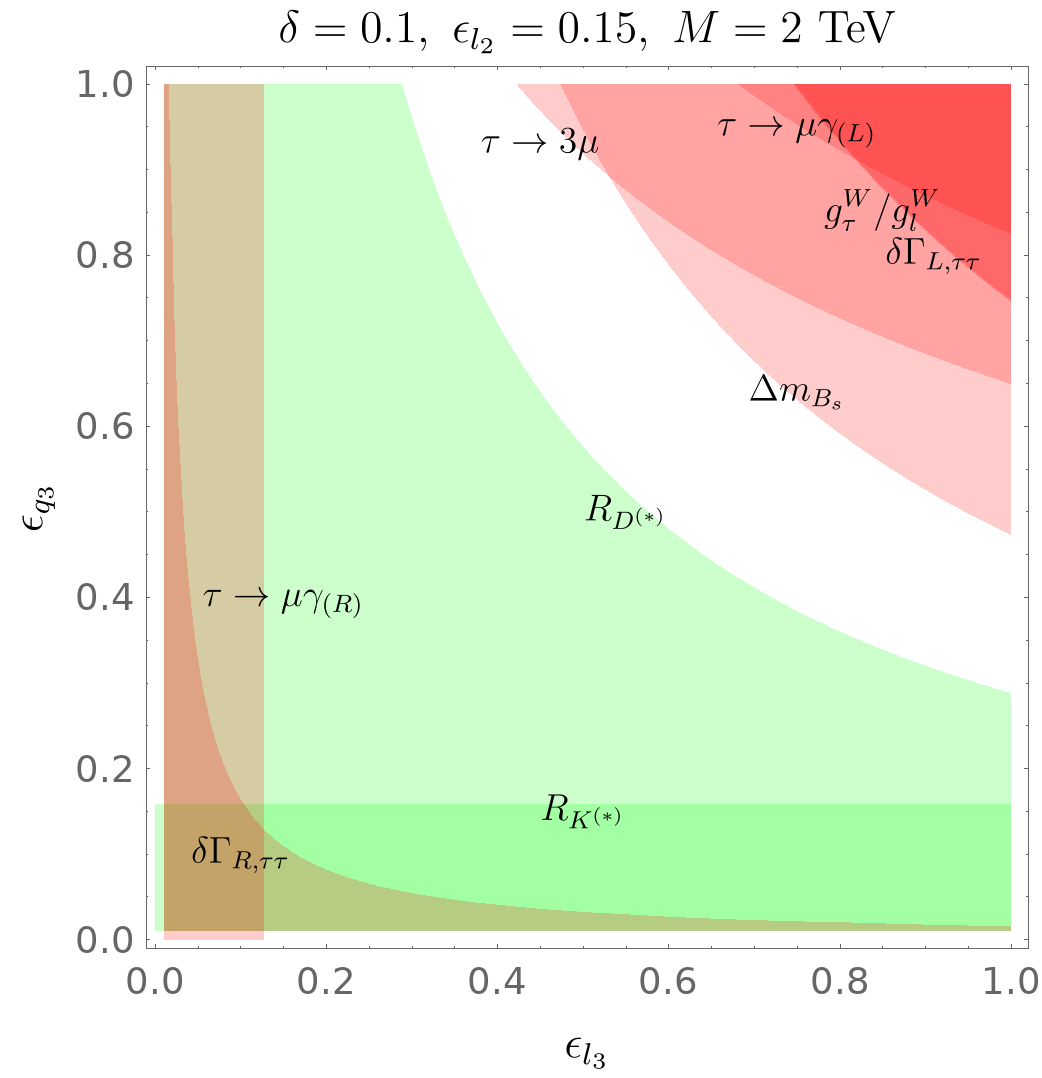}
  \includegraphics[width=0.495\textwidth]{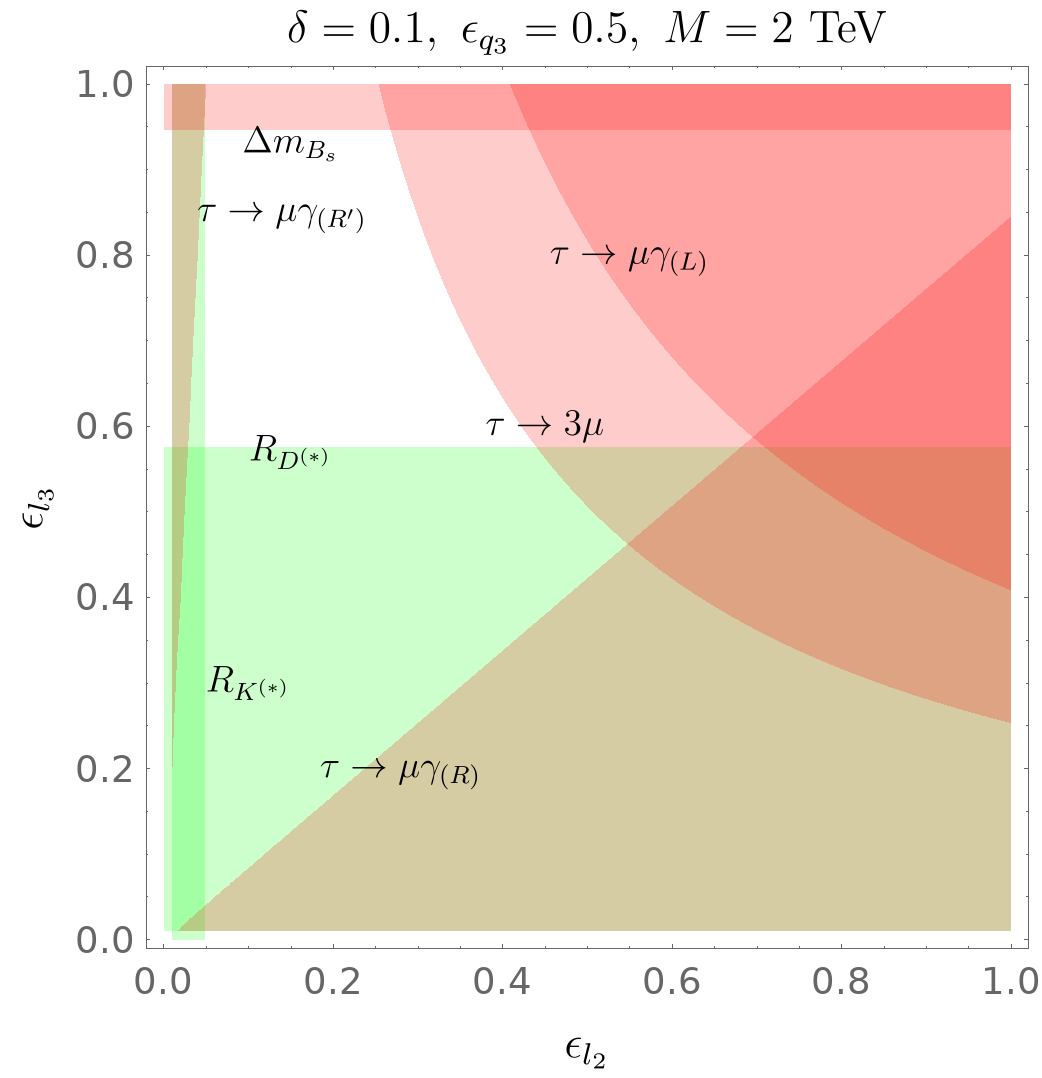} \\
 \includegraphics[width=0.495\textwidth]{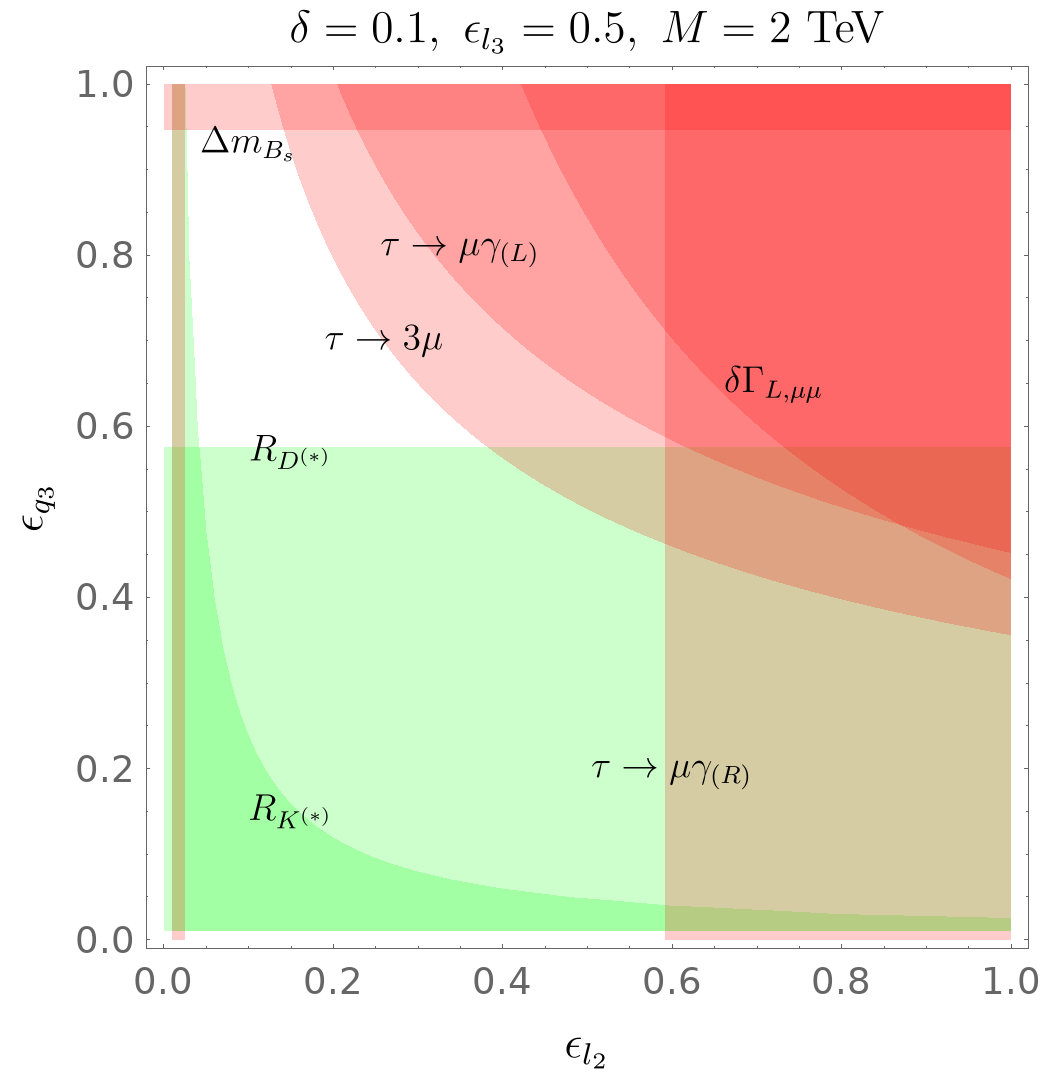}
  \caption{Scans in each pair of compositeness fraction $\epsilon_f$, for fixed $\d1=$ values.}
  \label{3cuts}
\end{figure}
In Fig.~\ref{3cuts} we fix $\d1 = 0.1$ and $M = 2\ \rm{TeV}$, that are of expected values according to the estimates of sec.~\ref{sec-model}.  We also fix in each case one of the compositeness fractions $\epsilon_f$, and scan along the other two. We include all three of those sections for a better picturing of this dependence. Looking at the first section of Fig.~\ref{3cuts}, in the plane $\epsilon_{l3} - \epsilon_{q3}$, we see that the allowed region is limited by $\tau \to 3 \mu$, $\Delta m_{B_s}$ and $R_{D^{(*)}}$. As can be seen from Eqs.~(\ref{tau3m:eq}),~(\ref{Dmb:eq}) and~(\ref{rd:eq}), the window moves with $\d1$ and powers of $\epsilon_f$. As $\tau \to 3 \mu$ depends quadratically on $\d1$, this limit moves faster with increasing $\d1$ than the others. As the dependence is on positive powers of these $\epsilon_f$, an increase in $\d1$ will translate into a decrease of the allowed values of these coefficients, thus lowering the location of the window. The remaining fixed parameters in this figure are $\epsilon_{l2}$ and $M$. Although not all the quantities depend explicitly on the LQ mass $M$, there are those that do in different ways. For example, Eq.~(\ref{Dmb:eq}) shows that $\Delta m_{B_s}$ depends quadratically on $M$, whereas $\tau\to\mu\gamma(\rm{R})$ has an $M^{-4}$ dependence shown in Eq.~(\ref{taumgR:eq}). This means that the same figure, with a smaller value for $M$, will have a less stringent bound imposed by $\Delta m_{B_s}$, but a much more restrictive bound imposed by $\tau \to \mu \gamma (\rm{R})$. The dependence on $\epsilon_{l2}$ can be seen by looking at the other sections in Fig.~\ref{3cuts}, or by looking at the expressions above. For example, as $R_{K^{(*)}}$ depends on $\epsilon_{l2}$, we see how a lower value of $\epsilon_{l2}$ will make the bound imposed by $R_K$ on the minimum $\epsilon_{q3}$ to increase, eventually becoming one of the bounds on the allowed window. The same reasoning can be applied to the other sections on the figure. In the plane $\epsilon_{l2} - \epsilon_{l3}$ we can see the two limits imposed by the two contributions to $\tau \to \mu \gamma (\rm{R})$, where one dominates for $\epsilon_{l2} \gtrsim \epsilon_{l3}$, and the other in the limit $\epsilon_{l3} \gg \epsilon_{l2}$. These bounds are not as relevant for $M=2$ TeV, however, but decreasing the value of the LQ mass to 1 TeV makes them become two of the most important bounds for the allowed window, surpassing the limits imposed by $R_{K^{(*)}}$ and by $\tau \to 3 \mu$.

\begin{figure}[h!]
  \centering
  \includegraphics[width=0.495\textwidth]{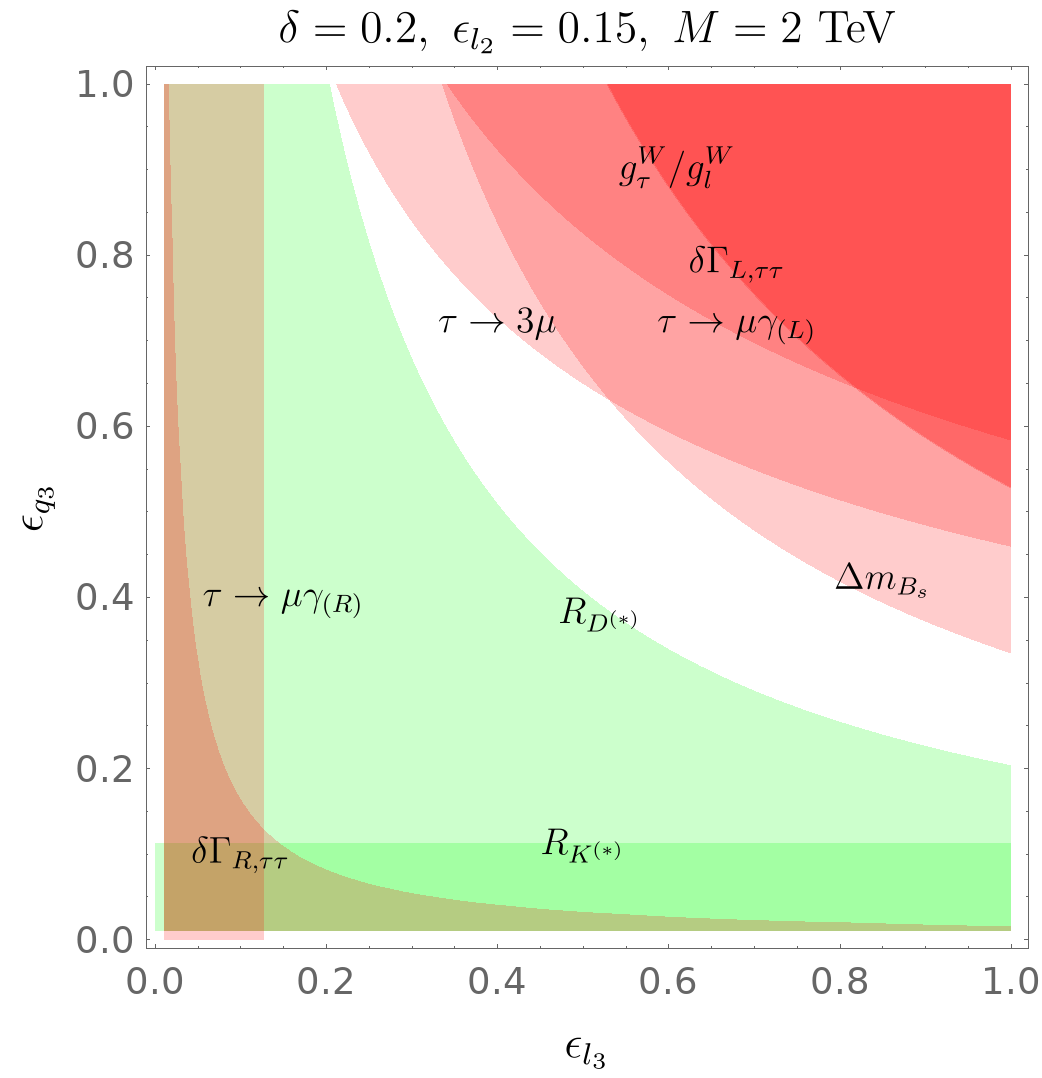}
  \includegraphics[width=0.495\textwidth]{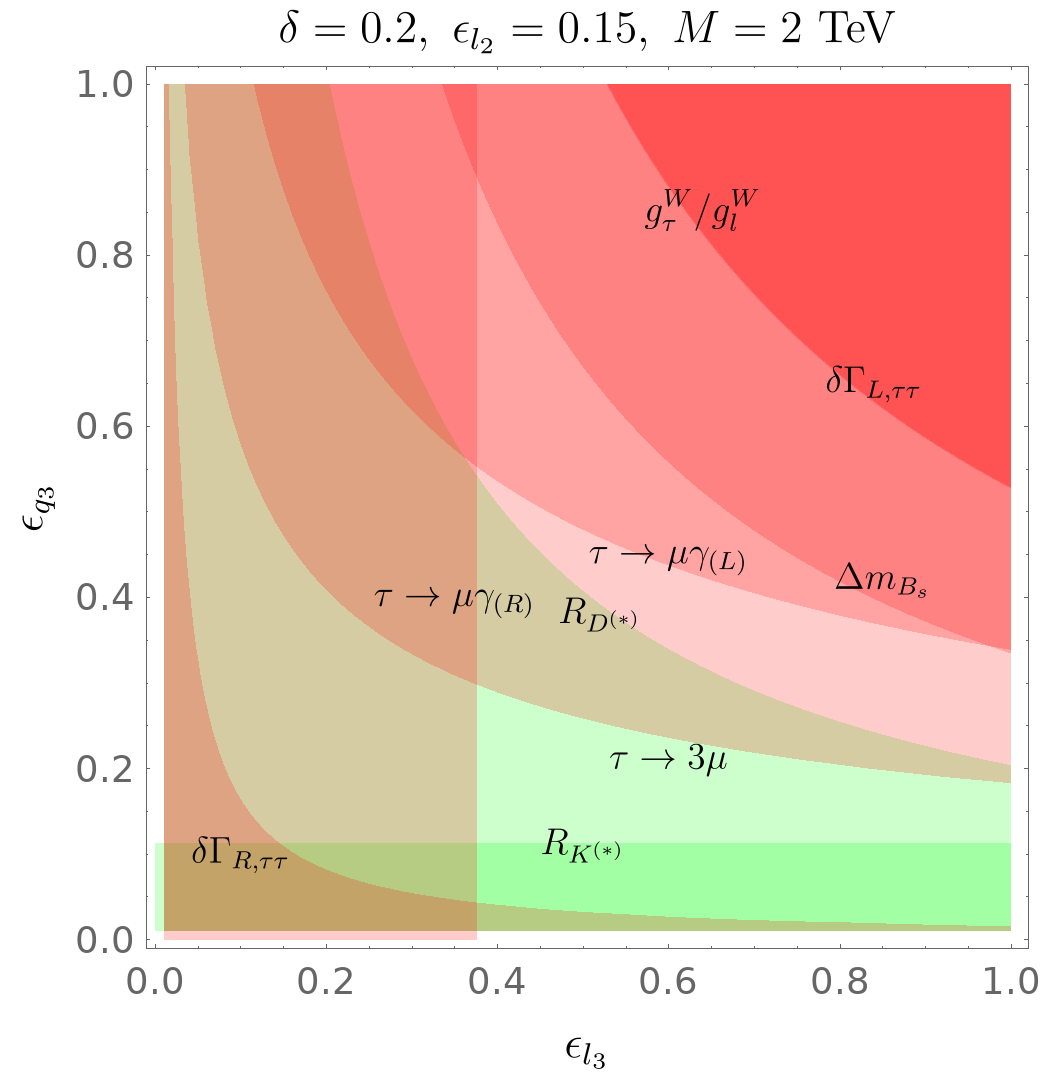}
  \caption{Left panel: Current bounds, Right panel: expected increase in sensitivity for LFV in $\tau$ decays} 
  \label{BelleII}
  \end{figure}
The bounds will change in the future, as the precision of experiments improves, particularly interesting is $\tau \to 3 \mu$. For instance, in Belle II the expected sensitivity for the branching ratios in LFV searches in $\tau$ decays improves by either one or two orders of magnitude~\cite{1808.10567}. We expect:
\al
   {\rm Br}(\tau \to \mu \gamma) &= 4.4\times10^{-8} \to 5 \times 10^{-9} \ \quad {(90\%{\rm CL})},\nn \\
   {\rm Br}(\tau \to 3 \mu) &=  1.2\times10^{-8}  \to 3 \times10^{-10} \ \quad {(90\%{\rm CL})}.
\fal
We can then check how the new bounds look on our 2d scans in $\epsilon_f$ space. For example, for $\epsilon_{l2} = 0.15$, we show the current and the expected bounds, side by side in Fig.~\ref{BelleII}. There we use a different value for one of the parameters, compared with Fig.~\ref{3cuts}: $\d1 =0.2$. In the left we show how some of the curves get modified by the enlargement of $\d1$, whereas on the right we show the expected increase in sensitivity. The limit imposed by $\tau \to 3 \mu$ rules out the selected window, meaning that either a higher tuning would be needed to pass the constraints, or that some violation of this quantity would have to be observed. We can tune some parameters to recover the window, for instance by lowering $\epsilon_{l2} =0.08$ and increasing $M=3$ TeV, we get a small window for $\epsilon_{l3} \simeq 1$. In this case the window is small and is located around $\epsilon_{q3}\simeq 0.3$, a somewhat low degree of compositeness compared with the usual scenarios of composite Higgs models.

\section{Spectrum of resonances}\label{sec-resonances}
In this section we describe the phenomenology of the composite model, focusing on the spectrum of resonances, both spin 1 and 1/2. The scale of the masses of these resonances is $m_* \simeq g_* f \sim 10-30\ \rm{TeV}$. The quantum numbers of the resonances are set by the group theory alone in the case of the spin-1 resonances, or by the embeddings of the SM fermions in irreducible representations of the global symmetry group of the SCFT.

{{Before describing those resonances we analyse very briefly the LHC phenomenology of the spin 0 states. Pair production of $S_1$ and $S_3$ by QCD interactions depend only on the LQ masses to leading order, whereas single production is more model dependent, being subleading for masses below $\sim 1.1-1.5$~TeV~\cite{Isidori}. Given the flavor structure of the couplings to SM fermions, the LQs decay predominantly to fermions of the third generation, moreover, the charge -1/3 states decay to $b\nu$ and $t\tau$ with similar branching fractions~\cite{1808.02063}. ATLAS~\cite{2101.11582,2101.12527} and CMS~\cite{2012.04178} have searched for these LQs in different final states, CMS taking into account contributions from double and single production, in the last case with couplings of order 1.5-2.5, that are of similar size as the couplings expected in the present model. Those analysis exclude masses below $\sim 1.1-1.2$~TeV, leading to the most stringent bounds today from direct searches. Although in our model some bounds from flavor physics require masses above the limits from LHC: $m_{\rm LQ}\gtrsim 1.5$~TeV, there are two LQs with charge -1/3 that could add and give a larger cross section than in the case of just one state, perhaps strengthening the bounds. This interesting situation deserves a careful analysis that is beyond the scope of this paper.}}

\subsection{Spin-1 resonances}
To obtain the quantum numbers of the spin-1 resonances we use that $\rm{\bf Adj}[\so{10}\tim\so{5}]= \left(\rm{\bf Adj}[\so{10}],{\bf 1}\right) \oplus \left({\bf 1},\rm{\bf Adj}[\so{5}]\right)
$, and we decompose these adjoint representations under the SM symmetry group. Regarding the $\so{5}$ adjoint representation, it is as in the MHCM, leading to resonances transforming under G$_{\rm SM}$ as:
\begin{equation}
{\bf \left(1,3 \right)_{\rm 0} + \left(1,1\right)_{\lbrace \rm +1,0,-1\rbrace} + (1,2)_{\pm \rm \frac12}}
\end{equation}
Thus before EWSB there are multiplets that transform as the $W$s and $B$ of the SM, along with new states transforming as charged weak doublet and singlet. After EWSB, we get states with charges $\pm 1$ and 0, similar to heavy resonances of $W$ and $Z$ bosons. 

When looking at the $\so{10}$ adjoint representation, using Eqs.~(\ref{eq-45-dec}), (\ref{eq-6-dec}) and (\ref{eq-repsSU4}), we get the following representations under G$_{\rm SM}$ for the remaining vectorial resonances:
\begin{equation}
{\bf (1,3)_{\rm 0} + (3,1)_{\rm -\frac13} + (3,3)_{\rm -\frac13} + (1,1)_{\rm 0} + (3,1)_{\rm \frac23} + (8,1)_{\rm 0} + (\rm{h.c.})} \ .
\end{equation} 
Here we recognize $W$-like, $Z$-like and gluon-like resonances, along with three representations transforming as color triplets and charged. If we look at their quantum numbers, we can identify them with LQs as:
\begin{align}
&{\bf (3,1)_{\rm -\frac13} }\to \bar{U}_1 \ ,\nn \\
&{\bf (3,1)_{\rm \frac23}} \to U_1 \ ,\nn \\
&{\bf (3,3)_{\rm -\frac13}} \to X \ .
\end{align} 
The state transforming as ${\bf (3,3)}_{\rm -1/3}$ which we call $X$, does not couple to the SM fermions through $d=4$ operators, and hence all possible interactions will be suppressed by powers of a higher scale. At the same order the LQ $\bar{U}_1$ only has interactions involving Right-handed neutrinos $\nu_R$, whereas $U_1$ has coupling with the doublets $q_L$ and $l_L$. However, if we look at the $\su{2}_A\tim\su{2}_B$ structure of the representations, we see there is no way to couple $U_1$ to both $q_L$ and $l_L$ at tree level, without further insertions of fields. This is because, under $\su{2}_A\tim\su{2}_B$, $U_1 \sim {\bf (1,1)}$, whereas $q_L \sim {\bf (2,1)}$ and $l_L \sim {\bf (1,2)}$, hence there is no singlet combination when multiplying these three fields.

We consider now the decay of these LQs. They are embedded in two different representations of $\so{6}$: $U_1$ is in ${\bf15}$, whereas $\bar{U}_1$ and $X$ are in the ${\bf 6}$. The lowest dimensional operator respecting the H symmetry that allow the decay of $U_1$ require one insertion of an scalar LQ. Using that for SO(6): ${\bf 4}\tim{\bf4}\tim{\bf 15}\tim{\bf 6}\supset{\bf 1}$, the following operators can be considered:
\begin{align}
\mathcal{O}^{6}_{U} = \left(\bar{q}^c_L  \sigma_{\mu\nu}  l_L\right) \, S_1 \, \partial^{[\mu}U_1^{\nu]} \ ,\qquad 
\tilde{\mathcal{O}}^{6}_{U} = \left(\bar{q}^c_L  \sigma_{\mu\nu} \sigma^a  l_L\right) \, S_3^a \, \partial^{[\mu}U_1^{\nu]}  \ .
\end{align} 
The decay into SM particles proceeds then through a scalar LQ, with a final state containing 4 SM fermions: $U_1\to \bar q\bar\ell S_{1,3}^\dagger\to\bar q\bar q'\bar\ell\bar\ell'$.

Regarding the LQs present in representation ${\bf (6,2,2)}$, we can write dimension 5 operators:
\begin{align}
\mathcal{O}^5_{\bar{U}} = \left(\bar{q}^c_L  \sigma_{\mu\nu}  l_L\right) \, \partial^{[\mu}\bar{U}_1^{\nu]*} \ , \qquad
\mathcal{O}^5_{X} = \left(\bar{q}^c_L  \sigma_{\mu\nu} \sigma^a l_L\right) \, \partial^{[\mu}X_a^{\nu]*} \ .
\label{eq:O5}
\end{align}
These two other states decay into a quark and a lepton, without a scalar LQ insertion. 

The WCs of these operators are expected to be generated at loop level, requiring also insertions of the mixing factors $\epsilon_q \, \epsilon_l$, that are dominated by those of the third generation, thus the final fermions are preferentially of the third generation.

{We can investigate the effect of these operators on the phenomenology by integrating-out the spin-1 states and estimating the size of their contributions to other WCs. Starting from the Lagrangian for a massive $\bar{U}_1$ and the interactions given by $ c_{\bar{U}} \mathcal{O}^5_{\bar{U}}$, at low energies we get the effective dimension 8 interaction
\be
\mathcal{L}_{qlql} = - \frac32 \frac{c^2_{\bar{U}}}{M^2_{\bar{U}_1}}  \partial_\mu(\bar{q}_L^c\sigma^{\mu \nu}l_L)\partial^\rho (\bar{q}_L^c \sigma_{\rho \nu} l_L)
\te
that is expected to be suppressed compared with the effect of dimension 6 operators. To estimate its effect on meson physics one has to make use of Fierz identities, to transform the Lorentz structure into the more familiar $(\bar{q}_L M q) (\bar{l}_L M l_L)$. This is a somewhat involved process, as the matrix structure is not the usual $\sigma_{\mu \nu} \sigma^{\mu \nu}$ one, this one having two free Lorentz indices that are contracted with derivatives. Although the analysis of dimension 8 operators is beyond the scope of this work, one can make an estimate of the size of their WCs assuming that the energies are of order GeV, obtaining a coefficient:
$
(\partial)^2 c_{\bar{U}_1}^2/M_*^2 \sim ({\rm GeV})^2 c_{\bar{U}_1}^2/M_*^2
\label{wcd8}
$.
This WC can be compared with those generated by the scalar LQs for dimension 6 operators at tree level, that are of order $\sim x^2/M_{1,3}^2$. Assuming that $c_{\bar{U}}$ is generated at loop level: $c_{\bar{U}} \sim (g_*/4 \pi)^2 g_* \epsilon_q \epsilon_l/M_*$, one can estimate the ratio to be 
\be
\left(\frac{g_*}{4\pi}\right)^4 \left(\frac{{\rm GeV}}{M_*}\right)^2 \left( \frac{M_{1,3}}{M_*}\right)^2 \sim 1.06 \times 10^{-12} \left(\frac{g_*}{4}\right)^4 \left(\frac{M_{1,3}}{{\rm TeV}}\right)^2 \left( \frac{10\ {\rm TeV}}{M_*}\right)^4
\te
If the operator $\mathcal{O}^5_{\bar{U}}$ were generated at tree-level, then this ratio would be enhanced by a factor $(4\pi/g_*)^4$, giving a ratio of $\sim 10^{-10}$.
We therefore can expect the effect of the vector LQs on the meson phenomenology to be suppressed, since their interactions with the SM fermions arise from operators of dimension 5 or 6. 
}

\subsection{Fermionic resonances}
In the case of fermion fields one can proceed in a similar way to study their quantum numbers, decomposing their representations under G$_{\rm SM}$.

For ${\bf (16,5)}$ we obtain:
\begin{align}
{\bf (16,5)} \supset \bf (3,2)_{\rm \frac16} + (1,2)_{\rm -\frac12} + (3,1)_{\rm \lbrace \frac23,-\frac13\rbrace} + (3,3)_{\rm \lbrace \frac23,-\frac13\rbrace} 
+ {\bf (1,1)_{\rm \lbrace0,-1\rbrace} + (1,3)_{\rm \lbrace0,-1\rbrace} + \rm{h.c.}}
\end{align} 
leading to massive resonances with the same quantum numbers as the SM fields: $q_L$, $l_L$, $u_R$, $d_R$, $\ell_R$, as well as a singlet. Besides these states, we find fields similar to $u_R$, $d_R$, $\ell_R$ and $\nu_R$, with the exception that they transform as triplets under $\su{2}_L$, instead of singlets. This gives rise, after EWSB, to states with exotic charges, the color triplets with $Q=5/3,-4/3$, and the color singlet with $Q=-2$. 

When decomposing ${\bf (\overline{144},5)}$ under the SM group we find a set of fields having, under G$_{\rm SM}$, the same properties appearing in ${\bf (\overline{16},5)}$. Besides them, with the appearance of representations ${\bf (4,3,2)}$ and ${\bf (\bar{4},2,3)}$, we get similar states but forming different multiplets under $\su{2}_L$. For example, we get a quark-like state with quantum numbers ${\bf(3,4)}_{1/6}$, along with other states that transform as singlets, triplets and quintuplets under the weak group. Finally, when observing the representations that come from the decomposition of the ${\bf 20}$ of $\so{6}$, we get:
\begin{align}
&\bf (3,1)_{\rm \lbrace \frac23,- \frac13\rbrace} + (3,2)_{\rm \frac16} + (3,3)_{\rm \lbrace \frac23,-\frac13\rbrace} + (\bar{3},1)_{\rm \lbrace \frac43 , \frac13 \rbrace} + (\bar{3},2)_{\rm \frac56} + (\bar{3},3)_{\rm \lbrace \frac43 , \frac13 \rbrace} \nn \\
&\bf + (\bar{6},1)_{\rm \lbrace \frac23, -\frac13\rbrace} + (\bar{6},2)_{\rm \frac16} + (\bar{6},3)_{\rm \lbrace \frac23, -\frac13\rbrace} + (8,1)_{\rm \lbrace 0, -1\rbrace} + (8,2)_{\rm -\frac12} + (8,3)_{\rm \lbrace 0, -1\rbrace}
\end{align}
The $\bf{\overline{20}}$ contains the conjugate representations, that, besides the aforementioned states, leads to a new exotic color triplet with $Q=-7/3$. This state decays into another exotic-charged state of $Q=-4/3$, which then decays into SM states. In addition we find other states that transform as a sextet of $\su{3}_c$. Given the algebra of $\su{3}$: ${\bf \bar{3}}\tim{\bf\bar{3}} = {\bf3} + \bar{\bf 6}$, a $\bar{\bf6}_{2/3}$ decays into two color anti-triplets: a scalar LQ and a SM quark, leading to a final state with two quarks and one lepton after the decay of the scalar LQ. Notice that these interactions are allowed by the $\so{6}\tim\su{2}_A\tim\su{2}_B$ subgroup of $\so{10}$, since ${\bf20}\tim{\bf6}\tim{\bar{\bf4}}\supset{\bf 1}$, thus an invariant can be formed with a resonance in a sextet, one LQ and one SM antiquark. The treatment for the octet is similar, it decays through an intermediate scalar LQ. The octet with $Q=-2$ decays through a LQ of charge $-4/3$ and an anti-top.

\section{Conclusions}
We have proposed a model to explain the $B$-anomalies, investigating its capacity to simultaneously pass the bounds from other flavor observables. We have considered a strongly coupled theory based on a global symmetry group $\so{10}\tim\so{5}$, spontaneously broken to $\so{6}\tim\su{2}_A\tim\su{2}_B\tim\so{4}$ by the strong dynamics. This pattern of symmetries have several properties:  it contains the SM gauge symmetry group, it develops only the LQs $S_1$ and $S_3$, and the Higgs, as NGBs, it contains a custodial symmetry. We have determined the embeddings of the SM fermions into the larger symmetry group, selecting by phenomenological reasons ({\bf 16,5}) and $({\bf \overline{144},5})$, as well as their conjugates. 
We have shown that the embedding of all the fermions in ({\bf 16,5}) and its conjugate results in Left-handed LQ interactions that are equal for $S_1$ and $S_3$, thus they can not accommodate simultaneously the flavor constraints and the $B$-anomalies. We have shown that mixing the lepton doublet with a ${\bf \overline{144}}\tim{\bf 5}$ can solve this problem. Moreover, mixing it with resonances in both representations allows for couplings with $S_1$ and $S_3$ that are independent.
We have considered an anarchic flavor structure of the SCFT that, along with partial compositeness, give a rationale for the SM fermion spectrum and mixings, and contains a GIM-like mechanism suppressing flavor transitions. As is well known, this flavor framework does not pass some bounds from meson mixing, thus we have assumed a scale of resonances of order 10-30~TeV, increasing the amount of tuning required for the EW scale, that is estimated to be at least of order $0.1-1\%$.

We have considered an effective description of the dynamics where only the NGBs and the SM fields are kept, armed with it we have shown how to compute the one loop potential, estimating the masses of the leptoquarks in the range of few TeV. We have also computed the Higgs potential, that is similar to the MCHM based on SO(5)/SO(4). Besides, we have estimated the corrections of the heavy resonances to the $Z b_L \bar b_L$ coupling, which, due to the large degree of compositeness of the third generation quarks, gets corrections that are near the saturation of the bound. This signals that certain amount of tuning could be required for this observable. We have also discussed briefly the proton decay, that is forbidden by a discrete symmetry.

We have estimated the size of the contributions of the scalar LQs to the $B$-anomalies and to the flavor observables that pose the most stringent constraints, some of these contributions arise at tree level and others at loop level. For that analysis we have used the hypothesis of anarchic partial compositeness.
We have performed scans in the degrees of compositeness of second and third generation of leptons, the third generation of quarks, the masses of the LQs and the strength of the coupling between composite resonances. We found that a tension arises between an explanation of $R_{D^{(*)}}$ and some flavor observables, mostly $\tau \to 3 \mu$, but also $Z\nu\bar\nu$ and $\tau \to \mu \gamma$, that requires a tuning of order $10-25\%$.
We have defined a window in parameter space with ``minimal tuning'', this window requires sizeable degrees of compositeness for third generation $l_L$ and $q_L$, but the amount of compositeness is also bounded from above by some flavor constraints, particularly $\tau \to 3 \mu$ and $\Delta m_{B_s}$. We have shown how some of these flavor constraints are expected to change in the future, introducing even more tension with $R_{D^{(*)}}$, and have shown the change in this window accordingly. {We have also considered observables $\mu \to 3 e$ and $\mu \to e \gamma$. We found that the former can be easily accommodated by our model, while the latter comes into conflict with the expected degree of compositeness for the electron. This can be solved with the introduction of small bilinear couplings, that for the first generation allow to decouple its mass from its degree of compositeness.}

Several authors have considered the possibility to explain also the anomalous magnetic moment of the muon with the presence of scalar leptoquarks. In anarchic partial compositeness the estimate for the correction to this quantity is independent of the fermion degree of compositeness, depending only on the mass of the LQs. An explanation of the experimental result would require a rather small LQ mass, $M \lesssim 250$ GeV, incompatible with direct search bounds, or a higher amount of tuning in the anarchic coefficients. 


We have analysed the spectrum of resonances, finding heavier copies of SM particles, as well as exotic states. Regarding vector resonances, we have found resonances of the $W$ and $Z$ bosons, as well as heavy gluons, plus three colored states that can be associated with leptoquarks. However, none of these leptoquarks can couple to SM fermions with $d=4$ operators, either because of their quantum numbers, or because of the $\su{2}_A\tim\su{2}_B$ symmetry. We have shown the smallest dimensional operators that allow these leptoquarks to decay into SM particles. 
Regarding fermionic the resonances, besides the states with the same charges as the SM ones, there are exotic states with charges $-7/3,\ -4/3,\ 5/3$, that are color triplets or sextets, as well as color octets and singlets with integer charges. 

Finally let us comment on a few possible directions that could be investigated. We have estimated many quantities assuming generic properties of the theory of resonances, it would be interesting to compute them by considering specific realizations, as discrete composite models, or extra-dimensions. On a different direction, since some of the leading constraints are related with modifications of $Z$-couplings, it would be interesting to explore other representations of fermions that could protect them, and eventually relax the tension between the anomalies and some of the flavor constraints.

\section*{Acknowledgments}
We thank David Marzocca for clarifications on some results of Ref.~\cite{Isidori}. We also thank the referee for pointing to us that $\mu \to e \gamma$ could be a problem for the model. This project has been partially supported by CONICET Argentina with PIP-0299 and FONCyT with PICT-2018-03682.


\appendix

\section{Embeddings of $l_L$ and LQ couplings}\label{ap-cg}
Putting $q_L$ in ${\bf (4,2,1)}$ and $l_L$ in ${\bf (4,1,2)}$ is problematic when constructing the interaction term for the LQs, because the Lagrangian couples $S_1$ and $S_3$ to $q_L l_L$ with equal strength. That is in equation \ref{eq-LS1}, we get that $\xs{} = \xt{}$. This alignment is insufficient when trying to explain the B anomalies: for example, it gives no correction to $R_{D^{(*)}}$, as can be seen in Eq. (\ref{rd:eq}).

To solve this problem we have to consider different fermion representations. First we have to understand why we get the same couplings for $S_1$ and $S_3$ when using the representations above. It is enough to look at the $\su{2}_A\tim \su{2}_B$ representations, as the color contraction is straightforward between $q$ and $S$, and regarding $\so{5}$ they are all in trivial representations.

In the scheme previously defined, we have the LQ belonging in a bidoublet $S_{\a \b} \sim ({\bf 2, 2})$, whereas quark and lepton are embedded in a single doublet: $q_{\a'} \sim {\bf (2,1)}$, $l_{a,\b'} \sim {\bf (1,2)}$. To write an invariant we start with the combination 
\be
S_{\a \b}\, q_{\a'}\, l_{a,\b'} \,G_{\a \b \a' \b'} \ .\nn
\te 
We use the following Clebsch-Gordan coefficients for ${\bf 2}\tim{\bf 2} \to {\bf 1} + {\bf 3}$,
\al
C_{\a\a'}^0 &= \frac{\delta_{\a \ua} \delta_{\a'\da} - \delta_{\a \da} \delta_{\a'\ua} }{\sqrt{2}} \ , \nn \\
C_{\a\a'}^{1,k} &= \delta_{k,1} \delta_{\a \ua} \delta_{\a' \ua} +  \delta_{k,0} \frac{\delta_{\a \ua} \delta_{\a'\da} + \delta_{\a \da} \delta_{\a'\ua} }{\sqrt{2}} + \delta_{k,-1} \delta_{\a\da} \delta_{\a\da} \ ,\nn
\fal
(where we represent spin half with up and down arrows, and integer spin with the integer $k$). As we are combining doublets, the invariant combination we have $G_{\a \b \a' \b'} = C^0_{\a \a'} C_{\b \b'}^0$. Replacing in the above formula, 
\be
S_{\a \b}\, q_{\a'}\, l_{a,\b'} \,G_{\a \b \a' \b'} = \frac12 \left( S_{\ua \ua} l_{a,\da} q_\da + S_{\da \da} l_{a,\ua} q_\ua - S_{\ua \da} l_{a,\ua} q_\da - S_{\da\ua} l_{a,\da} q_\ua  \right) \ .\nn
\te
We can rewrite these LQ states in terms of the triplet and the singlet:
\al
S_{\ua \ua} = S_3^1\ , 
&\quad\quad 
S_{\da \da}= S_3^{-1} \ ,\nn \\
S_{\ua \da} = \frac{S_3^0 + S_1}{\sqrt{2}} \ ,
&\quad\quad 
S_{\da \ua} = \frac{S_3^0 - S_1}{\sqrt{2}} \ .\nn
\fal
By doing this we get (omitting an overall factor of $\frac12$)
\be
S_3^1 l_{a,\da} q_\da + S_3^{-1} l_{a,\ua} q_\ua - S_3^0 \frac{l_{a,\ua} q_\da + l_{a,\da} q_\ua}{\sqrt{2}}  - S_1 \frac{l_{a,\ua} q_\da - l_{a,\da} q_\ua}{\sqrt{2}} \ .\nn
\te
And here we see the same size of coupling for the $\su{2}$ singlet and triplet.

Let us consider now a different embedding for the $l_L$ that can differentiate between $S_1$ and $S_3$ couplings. We start by considering the representation ${\bf(3,2)}$. We write $l_{b,k \b'}$ for the degrees of freedom of a field transforming in that representation. If we take the full representation to be ${\bf(4,3,2,1,1)}$, the dynamical degrees of freedom of this lepton doublet will be those of ${\bf 2}\in {\bf 3}\tim{\bf 2}$. 

Once again we construct an invariant using $q$, $l_b$ and $S$. We write
\be 
S_{\a \b}\, q_{\a'}\, l_{b,k \b'} \,\tilde{G}_{\a \b \a' \b' k} \ .\nn
\te
The way to combine these fields into an invariant is now by the use of the CG: ${\bf 2}\tim{\bf 2} \to {\bf 3}$. We also have to make use of a matrix corresponding to a $\pi$ rotation around the $y$ axis (which corresponds to the CG for  ${\bf 3}\tim{\bf 3} \to {\bf 1}$),
\be
R_{kk'} = \begin{pmatrix}
  0&0&1\\
  0&-1&0\\
  1&0&0\nn
\end{pmatrix}
\te
\noindent in order to correctly contract two triplets. Now, we write the combination as
\be
\tilde{G}_{\a\b\a'\b'k} = R_{kk'} \, C^{1,k}_{\a\a'} \, C^0_{\b\b'} \ .\nn
\te
By replacing these matrices we get
\al
&\left(l_{b,1 \b'} S_{\da \b} \, q_\da + l_{b,-1 \b'} S_{\ua \b} \, q_\ua - l_{b,0 \b'} \frac{S_{\ua \b} \, q_\da + S_{\da \b} \, q_\ua }{\sqrt{2}} \right)C^0_{\b\b'} \ .\nn 
\fal
Again taking away an overall factor of $\frac12$ we have
\al 
&S_{\ua \ua} \left(\sqrt{2} l_{b,-1 \da} q_\ua - l_{b,0 \da} q_\da \right)  
+ S_{\da \da} \left(l_{b,0 \ua} q_\ua - \sqrt{2} l_{b,1 \ua} q_\da \right) 
+ S_{\ua \da} \left(l_{b,0 \ua} q_\da - \sqrt{2} l_{b,-1 \ua} q_\ua \right) 
\nn \\&
+ S_{\da \ua} \left(\sqrt{2} l_{b,1 \da} q_\da -l_{b,0 \da} q_\ua \right) \ .\nn
\fal
We can drop out the fields $l_{b,-1 \da}$ and $l_{b,1 \ua}$ as they are the highest (and lowest) spin components of the fourplet of $\su{2}_L$, of spin $3/2$. Regarding the other components, we can use CG table to write
\al
&l_{b,1,\da} = \sqrt{\frac{2}{3}} l_\ua \ , 
&l_{b,0,\ua} = -\frac{1}{\sqrt{3}} l_\ua \ ,\nn\\
&l_{b,-1,\ua} = -\sqrt{\frac{2}{3}} l_\da \ ,
&l_{b,0,\da} = \frac{1}{\sqrt{3}} l_\da \ .\nn
\fal
Where we only turn on the dynamical d.o.f. belonging to the doublet of $\su{2}_L$. 
Using this and rewriting the LQ states in terms of $S_{1,3}$ we arrive to
\be
\frac{1}{\sqrt{3}} \left( -S_3^1 l_\da q_\da  - S_3^{-1} l_\ua q_\ua + S_3^0 \frac{l_\ua q_\da + l_\da q_\ua}{\sqrt{2}}  - 3 S_1 \frac{l_\ua q_\da - l_\da q_\ua}{\sqrt{2}}  \right) \ .\nn
\te
Comparing to the previous formula we find both a relative sign difference and a different weight for the couplings of $S_1$ and $S_3$. This extra representation then allows us to have independent couplings for each of the LQs. If we name the couplings for $l_a$ and $l_b$ as $x_{l_a}$ and $x_{l_b}$, we have: 
\al
x_3 \propto x_{l_a} - \frac{x_{l_b}}{\sqrt{3}} \ , \qquad\qquad
x_1 \propto x_{l_a} + \sqrt{3}\ x_{l_b} \ . 
\fal
Thus we see how two embeddings for the lepton doublet allow for independent couplings for $S_1$ and $S_3$ LQs.

\section{Potential}\label{ap-potential}
In this appendix we describe some of the details involved in the calculation of the pNGB potential. Starting from Eq.~(\ref{eq-potentialCW}), we have the definition of the Coleman Weinberg potential. We want to calculate the fermionic contributions to the effective potential, particularly of the scalar LQs. However, as the quantity $\log\det\ {\cal K}(\Pi)$ does not have in general a closed form, we must expand the potential in powers of the pNGB. Moreover, as this potential contains a constant divergent term, we regularize it by subtracting the same expression but evaluated at $\Pi=0$. Here we can make use of an operator identity
\be
\log\det\ {\cal K}(\Pi) - \log \det\ {\cal K}(0) = \Tr \log {\cal K}(\Pi) - \Tr \log {\cal K}(0) = \Tr \log{\cal K}(\Pi){\cal K}^{-1}(0) \ .
\te
In order to expand in powers of the pNGB fields, we can introduce a factor $\omega$ accompanying the scalar fields, $\Pi \to \omega \Pi$, and expand the matrix ${\cal K}$ in powers of $\omega$, at the end of the calculation we set $\omega = 1$:
\be
{\cal K}(\Pi) = \sum_{n\geq 0} \omega^n {\cal K}_n \ .
\te
In this manner one can expand the expression above in powers of $\omega$. As we are interested mainly in the quadratic terms, for the leptoquark masses, we can write:
\be
 \Tr \log\left( {\cal K}(\Pi){\cal K}^{-1}(0) \right) = \omega \  \Tr\left( \tilde{\cal K}_1 \right)+ \omega^2 \ \Tr \left(\tilde{\cal K}_2 - \tilde{\cal K}_1^2/2\right) + {\cal O}\left(\omega^3\right)
\te
where for briefness we have defined $\tilde{\cal K}_n \equiv {\cal K}_n {\cal K}_0^{-1}$.
In the same way one can write all the higher order interaction terms. This way, the problem of expanding the potential in powers of the pNGB fields reduces in expanding the effective lagrangian in Eq.~(\ref{eq-LeffV}), writing the corresponding matrices, and taking traces. The linear term in the potential is zero, because no invariant can be formed by a single field.

We choose a basis for writing these matrices: $\lbrace u_L^c, d_L^c, u_R^c, \ell_L , \nu_L \rbrace$, with $c$ being a color index, obtaining $11\times 11$ matrices. We choose these degrees of freedom because they have the largest mixing angles and thus the highest contribution to the potential.  

In the following we will change $\Pi_{ff'}^{{\bf r}_H}(p)\to\pslash\ \Pi_{ff'}^{{\bf r}_H}(p)$ for correlators involving elementary fermions with the same chirality. 

For the masses of the LQs, defined according to Eq.~(\ref{eq-MLQ1}), we get:
\begin{align}
\tilde M^2 = \int\frac{d^4p}{(2\pi)^4}&\left[ 
\frac{\C_{l_al_a}^{\rob} -\C_{l_al_a}^{\roab}}{Z_l + \C_{l_a l_a}^{\rob} + \C_{l_bl_b}^{\rfttoo} } + 3 \frac{ \C_{qq}^{\roa} - \C_{qq}^{\robb}}{Z_q + \C_{qq}^{\roa}} 
\right.
\nonumber \\
&\left.+
3 \frac{\C_{uu}^{\rfa} -\C_{uu}^{\rfbb}}{Z_u + \C_{uu}^{\rfa}}
+\frac{(\C_{ql_a}^{\rob} -\C_{ql_a}^{\roab})^2}{2(Z_l + \C_{l_al_a}^{\rob} + \C_{l_bl_b}^{\rfttoo})(Z_q + \C_{qq}^{\roa})}
\right]
\nonumber
\end{align}
\begin{align}
\Delta M_1^2 = \int\frac{d^4p}{(2\pi)^4}&\left[
\ \frac{33 \C_{l_b l_b}^{\rs} + 10 \C_{l_b l_b}^{\ro} + 5 \C_{l_b l_b}^{\rc} - 48 \C_{l_b l_b}^{\rfttoo}}{9[Z_l + \C_{l_al_a}^{\rob} + \C_{l_bl_b}^{\rfttoo} ]} 
\right.
\nonumber \\
&\left.+
\frac{(45-36\sqrt{5})(\C_{ql_b}^{\roab})^2+36\sqrt{5}\C_{ql_b}^{\roab}\C_{ql_a}^{\rob}}{72(Z_l + \C_{l_al_a}^{\rob} + \C_{l_bl_b}^{\rfttoo})(Z_q + \C_{qq}^{\roa})}
\right]
\nonumber
\end{align}
\begin{align}
\Delta M_3^2 = \int\frac{d^4p}{(2\pi)^4}&\left[
\ \frac{\C_{l_b l_b}^{\rs} + 10 \C_{l_b l_b}^{\ro} + 5 \C_{l_b l_b}^{\rc} - 16 \C_{l_b l_b}^{\rfttoo}}{Z_l + \C_{l_al_a}^{\rob} + \C_{l_bl_b}^{\rfttoo}} 
\right.
\nonumber \\
&\left.+
\frac{(5+12\sqrt{5})(\C_{ql_b}^{\roab})^2-12\sqrt{5}\C_{ql_b}^{\roab}\C_{ql_a}^{\rob}}{72(Z_l + \C_{l_al_a}^{\rob} + \C_{l_bl_b}^{\rfttoo})(Z_q + \C_{qq}^{\roa})}
\right] \ .
\label{eq-MLQAp}
\end{align}

When calculating the potential for the Higgs component that acquires a vev, the pNGB matrices can be calculated to all orders in this field. Hence, we can calculate the one loop potential to all orders in $v$. We can write the following quadratic and quartic coefficients in Eq.~(\ref{eq-potential-vev1}), as integrals of the fermionic correlators:
\begin{align}\label{eq-ab-integrals}
\alpha &=  \int \frac{d^4p}{(2\pi)^4}\Bigg\lbrace \frac{4(\Pi_{l_a}^{\re}+\Pi_{l_b}^{\reb}-\Pi_{l_a}^{\rl}-\Pi_{l_b}^{\rlb})}{ Z_\ell + \Pi_{l_a}^{\rl}+ \Pi_{l_b}^{\rlb}} - \frac{12 (\Pi_q^{\rq}- \Pi_q^{\ru})}{Z_q + \Pi_q^{\rq}}  \nn \\
&-\frac32 \frac{( \Pi_{qu}^{\rq} - \Pi_{qu}^{\ru})^2}{( Z_q + \Pi_q^{\rq}) (Z_u + \Pi_u^{\ru})}  +\frac32 \frac{(\Pi_u^{\rq} - \Pi_u^{\ru})}{(Z_u + \Pi_u^{\ru})} \Bigg\rbrace\nn \\
\beta &=  \int \frac{d^4p}{(2\pi)^4}  \Bigg\lbrace \frac{2
  (\Pi_{l_a}^{\re} + \Pi_{l_b}^{\reb} - \Pi_{l_a}^{\rl} - \Pi_{l_b}^{\rlb})^2}{(Z_\ell + \Pi_{l_a}^{\rl} + \Pi_{l_b}^{\rlb})^2} +   \frac{6 (\Pi_q^{\rq}- \Pi_q^{\ru})^2}{\left(Z_q + \Pi_q^{\rq}\right)^2}  \nn \\
&+ \frac{3}{16} \frac{(\Pi_u^{\rq}-\Pi_u^{\ru})^2}{(Z_u+\Pi_u^{\ru})^2}  + \frac{3}{16} \frac{(\Pi_{qu}^{\rq}-\Pi_{qu}^{\ru})^4}{(Z_q + \Pi_q^{\rq})^2(Z_u+\Pi_u^{\ru})^2} \nn \\
&- \frac38  \frac{(\Pi_{qu}^{\rq}-\Pi_{qu}^{\ru})^2 (\Pi_u^{\rq}-\Pi_u^{\ru})  }{(Z_q + \Pi_q^{\rq})(Z_u+\Pi_u^{\ru})^2} - \frac23  \frac{(\Pi_{qu}^{\rq}-\Pi_{qu}^{\ru})^2 (\Pi_q^{\ru}-\Pi_q^{\rq}) }{(Z_q + \Pi_q^{\rq})^2(Z_u+\Pi_u^{\ru})} \Bigg\rbrace \ .
\end{align}

\section{Group representations}
In this appendix we briefly comment on the representations used in the calculations above, and on how to construct some of those representations. In this work we use a group consisting of the product of two groups, $\so{10}$ and $\so{5}$. Regarding $\so{5}$, we concern ourselves with the fundamental and the adjoint representations, whereas, for the $\so{10}$ factor we also have spinorial representations ${\bf 16}$, ${\bf 144}$ and their conjugates.
The generators of an $\so{N}$ group in the fundamental representation can be parametrized in a simple fashion by a set of matrices $\lbrace \left(\mathcal{T}_{lm}\right)_{jk}$; $l<m$; $m = 2,\dots, N \rbrace$
\be
\left(\mathcal{T}_{lm}\right)_{jk} = i \left(\delta_{lj} \delta_{mk} - \delta_{lk} \delta_{mj} \right) \ .
\te 
The adjoint representation can be constructed from the structure constants, or also by using the generators of the algebra as a basis for the vector space. As $\so{N}$ has $N (N-1)/2$ generators, one defines a vector transforming in adjoint representation as a linear combination of said generators.

More interesting is how to build the spinorial representations ${\bf 16}$ and ${\bf \overline{16}}$. This can be achieved by constructing a $32$-dimensional Clifford algebra of matrices $\Gamma^a$, $a \in \lbrace 1 \dots 10\rbrace$. These $\Gamma$ matrices can be built by tensor products of 5 Pauli matrices. They follow a simple structure, as 
\begin{alignat}{2}
\Gamma_1 &= \sigma_2 \otimes \sigma_3  \otimes \sigma_3  \otimes \sigma_3  \otimes \sigma_3 &\ ,\quad\quad\quad   \Gamma_6 &= - \sigma_1 \otimes \sigma_3  \otimes \sigma_3  \otimes \sigma_3 \otimes \sigma_3 \ ,\nn \\
\Gamma_2 &=\mathbb{I} \otimes \sigma_2 \otimes \sigma_3  \otimes \sigma_3  \otimes \sigma_3 \ ,&  \Gamma_7 &=- \mathbb{I} \otimes  \sigma_1 \otimes \sigma_3  \otimes \sigma_3  \otimes \sigma_3 \ , \nn \\
&\dots &  &\dots \nn  \\ 
\Gamma_5 &= \mathbb{I} \otimes \mathbb{I} \otimes \mathbb{I} \otimes \mathbb{I} \otimes \sigma_2 \ , & \Gamma_{10} &=  -\mathbb{I} \otimes \mathbb{I} \otimes \mathbb{I} \otimes \mathbb{I} \otimes \sigma_1 \ . 
\end{alignat}
One also needs to define $\Gamma_{11} \equiv {\displaystyle (-i)^5 \prod_{a} \Gamma^a}$, which anticommutes the other 10 matrices. With these matrices one can build the generators in the spinorial representation by use of the commutators
\be
\Sigma_{ab} = \frac{i}{4} \left[ \Gamma^a, \Gamma^b\right] \ .
\te
Now this produces 32-by-32 matrices which we need to disentangle into representations ${\bf 16}$ and ${\bf \overline{16}}$. We can do this by noting that $\Gamma_{11}$ commutes with all generators, and its eigenvalues are $\pm 1$. Thus, by diagonalizing $\Gamma_{11}$, we get block diagonal generators $\Sigma_{ab}$ corresponding to both representations. \cite{Raby}

Finally, in order to have different $q_L l_L S_{1,3}$ couplings, we need to consider representation ${\bf \overline{144}}$. One way to construct this representation is by the multiplication of smaller representations. We find the following product is the smallest that contains this representation
\be
{\bf 16} \tim {\bf 10} \to {\bf \overline{144}} + {\bf \overline{16}}
\te
We start from these two representations, we have matrices $\{T_a^{(10)}\}$ and $\{T_a^{(16)}\}$. We construct the product representation of this algebra by taking the Kronecker product between these matrices and the identity matrix
\be
T_a^{(160)} = T_a^{(10)} \otimes {\bf 1}^{(16)} +   {\bf 1}^{(10)} \otimes T_a^{(16)}
\te
These matrices generate the algebra in a reducible representation of dimension 160. We need to split them into two blocks corresponding to irreducible representations (irreps) ${\bf \overline{144}}$ and ${\bf \overline{16}}$. This amounts to finding the two orthogonal subspaces corresponding to these irreps. One way of finding these subspaces, is by using the quadratic Casimir. It so happens that the eigenvalues of the quadratic Casimir of these two representations are distinct. Thus, we write this Casimir element, and then diagonalize it
\be
C_2 \equiv \sum_{a} T_a^{(160)} T_a^{(160)} \to U_c C_2 U_c^\dagger = C_2^{\rm diag}
\te
This unitary transformation is the one that defines the two orthogonal subspaces, and thus makes each of the generators split into the two blocks corresponding to each one of the irreps:
\be
U_c T_a^{(160)} U_c^\dagger = T_a^{(\overline{144})} \oplus T_a^{(\overline{16})}
\te
In this manner one can easily build the 144 dimensional representation of $\so{10}$

\bibliographystyle{JHEP}
\bibliography{biblio}

\providecommand{\href}[2]{#2}\begingroup\raggedright\begin{thebibliography}{10}

\bibitem{RD_Babar}
{\scshape BaBar} collaboration, J.~Lees et~al., \emph{{Measurement of an Excess
  of $\bar{B} \to D^{(*)}\tau^- \bar{\nu}_\tau$ Decays and Implications for
  Charged Higgs Bosons}},
  \href{http://dx.doi.org/10.1103/PhysRevD.88.072012}{\emph{Phys. Rev. D} {\bf
  88} (2013) 072012}, [\href{http://arxiv.org/abs/1303.0571}{{\tt 1303.0571}}].

\bibitem{RD_Belle}
{\scshape Belle} collaboration, S.~Hirose et~al., \emph{{Measurement of the
  $\tau$ lepton polarization and $R(D^*)$ in the decay $\bar{B} \to D^* \tau^-
  \bar{\nu}_\tau$}},
  \href{http://dx.doi.org/10.1103/PhysRevLett.118.211801}{\emph{Phys. Rev.
  Lett.} {\bf 118} (2017) 211801}, [\href{http://arxiv.org/abs/1612.00529}{{\tt
  1612.00529}}].

\bibitem{RD_LHCB}
{\scshape LHCb} collaboration, R.~Aaij et~al., \emph{{Measurement of the ratio
  of branching fractions $\mathcal{B}(\bar{B}^0 \to
  D^{*+}\tau^{-}\bar{\nu}_{\tau})/\mathcal{B}(\bar{B}^0 \to
  D^{*+}\mu^{-}\bar{\nu}_{\mu})$}},
  \href{http://dx.doi.org/10.1103/PhysRevLett.115.111803}{\emph{Phys. Rev.
  Lett.} {\bf 115} (2015) 111803}, [\href{http://arxiv.org/abs/1506.08614}{{\tt
  1506.08614}}]. [Erratum: Phys.Rev.Lett. 115, 159901 (2015)].

\bibitem{RK_LHCB}
{\scshape LHCb} collaboration, R.~Aaij et~al., \emph{{Test of lepton
  universality using $B^{+}\rightarrow K^{+}\ell^{+}\ell^{-}$ decays}},
  \href{http://dx.doi.org/10.1103/PhysRevLett.113.151601}{\emph{Phys. Rev.
  Lett.} {\bf 113} (2014) 151601}, [\href{http://arxiv.org/abs/1406.6482}{{\tt
  1406.6482}}].

\bibitem{RK_LHCB2}
{\scshape LHCb} collaboration, R.~Aaij et~al., \emph{{Test of lepton
  universality with $B^{0} \rightarrow K^{*0}\ell^{+}\ell^{-}$ decays}},
  \href{http://dx.doi.org/10.1007/JHEP08(2017)055}{\emph{JHEP} {\bf 08} (2017)
  055}, [\href{http://arxiv.org/abs/1705.05802}{{\tt 1705.05802}}].

\bibitem{RK_LHCB3}
{\scshape LHCb} collaboration, R.~Aaij et~al., \emph{{Search for
  lepton-universality violation in $B^+\to K^+\ell^+\ell^-$ decays}},
  \href{http://dx.doi.org/10.1103/PhysRevLett.122.191801}{\emph{Phys. Rev.
  Lett.} {\bf 122} (2019) 191801}, [\href{http://arxiv.org/abs/1903.09252}{{\tt
  1903.09252}}].

\bibitem{RK_Belle}
{\scshape Belle} collaboration, A.~Abdesselam et~al., \emph{{Test of lepton
  flavor universality in ${B\to K^\ast\ell^+\ell^-}$ decays at Belle}},
  \href{http://arxiv.org/abs/1904.02440}{{\tt 1904.02440}}.

\bibitem{EFT_Azatov}
A.~Azatov, D.~Bardhan, D.~Ghosh, F.~Sgarlata and E.~Venturini, \emph{{Anatomy
  of $b \to c \tau \nu$ anomalies}},
  \href{http://dx.doi.org/10.1007/JHEP11(2018)187}{\emph{JHEP} {\bf 11} (2018)
  187}, [\href{http://arxiv.org/abs/1805.03209}{{\tt 1805.03209}}].

\bibitem{EFT_Bhattacharya}
B.~Bhattacharya, A.~Datta, D.~London and S.~Shivashankara, \emph{{Simultaneous
  Explanation of the $R_K$ and $R(D^{(*)})$ Puzzles}},
  \href{http://dx.doi.org/10.1016/j.physletb.2015.02.011}{\emph{Phys. Lett. B}
  {\bf 742} (2015) 370--374}, [\href{http://arxiv.org/abs/1412.7164}{{\tt
  1412.7164}}].

\bibitem{EFT_Grinstein}
R.~Alonso, B.~Grinstein and J.~Martin~Camalich, \emph{{Lepton universality
  violation and lepton flavor conservation in $B$-meson decays}},
  \href{http://dx.doi.org/10.1007/JHEP10(2015)184}{\emph{JHEP} {\bf 10} (2015)
  184}, [\href{http://arxiv.org/abs/1505.05164}{{\tt 1505.05164}}].

\bibitem{EFT_Greljo}
A.~Greljo, G.~Isidori and D.~Marzocca, \emph{{On the breaking of Lepton Flavor
  Universality in B decays}},
  \href{http://dx.doi.org/10.1007/JHEP07(2015)142}{\emph{JHEP} {\bf 07} (2015)
  142}, [\href{http://arxiv.org/abs/1506.01705}{{\tt 1506.01705}}].

\bibitem{EFT_Calibbi}
L.~Calibbi, A.~Crivellin and T.~Ota, \emph{{Effective Field Theory Approach to
  $b\to s\ell\ell^{(')}$, $B\to K^{(*)}\nu\overline{\nu}$ and $B\to
  D^{(*)}\tau\nu$ with Third Generation Couplings}},
  \href{http://dx.doi.org/10.1103/PhysRevLett.115.181801}{\emph{Phys. Rev.
  Lett.} {\bf 115} (2015) 181801}, [\href{http://arxiv.org/abs/1506.02661}{{\tt
  1506.02661}}].

\bibitem{EFT_Bordone}
M.~Bordone, G.~Isidori and S.~Trifinopoulos, \emph{{Semileptonic $B$-physics
  anomalies: A general EFT analysis within $U(2)^n$ flavor symmetry}},
  \href{http://dx.doi.org/10.1103/PhysRevD.96.015038}{\emph{Phys. Rev. D} {\bf
  96} (2017) 015038}, [\href{http://arxiv.org/abs/1702.07238}{{\tt
  1702.07238}}].

\bibitem{Marzocca_1}
V.~Gherardi, D.~Marzocca and E.~Venturini, \emph{{Matching scalar leptoquarks
  to the SMEFT at one loop}},
  \href{http://dx.doi.org/10.1007/JHEP07(2020)225}{\emph{JHEP} {\bf 07} (2020)
  225}, [\href{http://arxiv.org/abs/2003.12525}{{\tt 2003.12525}}].

\bibitem{Marzocca_2}
V.~Gherardi, D.~Marzocca and E.~Venturini, \emph{{Low-energy phenomenology of
  scalar leptoquarks at one-loop accuracy}},
  \href{http://arxiv.org/abs/2008.09548}{{\tt 2008.09548}}.

\bibitem{Cata1}
M.~Bordone, O.~Cat\`a, T.~Feldmann and R.~Mandal, \emph{{Constraining flavour
  patterns of scalar leptoquarks in the effective field theory}},
  \href{http://arxiv.org/abs/2010.03297}{{\tt 2010.03297}}.

\bibitem{Cata2}
M.~Bordone, O.~Cat\`a and T.~Feldmann, \emph{{Effective Theory Approach to New
  Physics with Flavour: General Framework and a Leptoquark Example}},
  \href{http://dx.doi.org/10.1007/JHEP01(2020)067}{\emph{JHEP} {\bf 01} (2020)
  067}, [\href{http://arxiv.org/abs/1910.02641}{{\tt 1910.02641}}].

\bibitem{NP_Bauer}
M.~Bauer and M.~Neubert, \emph{{Minimal Leptoquark Explanation for the
  R$_{D^{(*)}}$ , R$_K$ , and $(g-2)_g$ Anomalies}},
  \href{http://dx.doi.org/10.1103/PhysRevLett.116.141802}{\emph{Phys. Rev.
  Lett.} {\bf 116} (2016) 141802}, [\href{http://arxiv.org/abs/1511.01900}{{\tt
  1511.01900}}].

\bibitem{NP_Fajfer}
S.~Fajfer and N.~Ko\v{s}nik, \emph{{Vector leptoquark resolution of $R_K$ and
  $R_{D^{(*)}}$ puzzles}},
  \href{http://dx.doi.org/10.1016/j.physletb.2016.02.018}{\emph{Phys. Lett. B}
  {\bf 755} (2016) 270--274}, [\href{http://arxiv.org/abs/1511.06024}{{\tt
  1511.06024}}].

\bibitem{NP_Barbieri}
R.~Barbieri, G.~Isidori, A.~Pattori and F.~Senia, \emph{{Anomalies in
  $B$-decays and $U(2)$ flavour symmetry}},
  \href{http://dx.doi.org/10.1140/epjc/s10052-016-3905-3}{\emph{Eur. Phys. J.
  C} {\bf 76} (2016) 67}, [\href{http://arxiv.org/abs/1512.01560}{{\tt
  1512.01560}}].

\bibitem{NP_Das}
D.~Das, C.~Hati, G.~Kumar and N.~Mahajan, \emph{{Towards a unified explanation
  of $R_{D^{(\ast)}}$, $R_{K}$ and $(g-2)_{\mu}$ anomalies in a left-right
  model with leptoquarks}},
  \href{http://dx.doi.org/10.1103/PhysRevD.94.055034}{\emph{Phys. Rev. D} {\bf
  94} (2016) 055034}, [\href{http://arxiv.org/abs/1605.06313}{{\tt
  1605.06313}}].

\bibitem{NP_Crivellin}
A.~Crivellin, D.~M\"uller and T.~Ota, \emph{{Simultaneous explanation of
  R(D$^{(*)}$) and $b \to s \mu^+ \mu^- $: the last scalar leptoquarks
  standing}}, \href{http://dx.doi.org/10.1007/JHEP09(2017)040}{\emph{JHEP} {\bf
  09} (2017) 040}, [\href{http://arxiv.org/abs/1703.09226}{{\tt 1703.09226}}].

\bibitem{NP_Becirevic}
D.~Be\v{c}irevi\'c, S.~Fajfer, N.~Ko\v{s}nik and O.~Sumensari,
  \emph{{Leptoquark model to explain the $B$-physics anomalies, $R_K$ and
  $R_D$}}, \href{http://dx.doi.org/10.1103/PhysRevD.94.115021}{\emph{Phys. Rev.
  D} {\bf 94} (2016) 115021}, [\href{http://arxiv.org/abs/1608.08501}{{\tt
  1608.08501}}].

\bibitem{NP_Boucenna}
S.~M. Boucenna, A.~Celis, J.~Fuentes-Martin, A.~Vicente and J.~Virto,
  \emph{{Phenomenology of an $SU(2) \times SU(2) \times U(1)$ model with
  lepton-flavour non-universality}},
  \href{http://dx.doi.org/10.1007/JHEP12(2016)059}{\emph{JHEP} {\bf 12} (2016)
  059}, [\href{http://arxiv.org/abs/1608.01349}{{\tt 1608.01349}}].

\bibitem{NP_Hiller}
G.~Hiller, D.~Loose and K.~Sch\"onwald, \emph{{Leptoquark Flavor Patterns
  \textbackslash{}\& B Decay Anomalies}},
  \href{http://dx.doi.org/10.1007/JHEP12(2016)027}{\emph{JHEP} {\bf 12} (2016)
  027}, [\href{http://arxiv.org/abs/1609.08895}{{\tt 1609.08895}}].

\bibitem{NP_Bhattacharya}
B.~Bhattacharya, A.~Datta, J.-P. Gu\'evin, D.~London and R.~Watanabe,
  \emph{{Simultaneous Explanation of the $R_K$ and $R_{D^{(*)}}$ Puzzles: a
  Model Analysis}},
  \href{http://dx.doi.org/10.1007/JHEP01(2017)015}{\emph{JHEP} {\bf 01} (2017)
  015}, [\href{http://arxiv.org/abs/1609.09078}{{\tt 1609.09078}}].

\bibitem{NP_Crivellin2}
A.~Crivellin, C.~Greub, D.~M\"uller and F.~Saturnino, \emph{{Scalar Leptoquarks
  in Leptonic Processes}},  \href{http://arxiv.org/abs/2010.06593}{{\tt
  2010.06593}}.

\bibitem{NP_Barbieri2}
R.~Barbieri, C.~W. Murphy and F.~Senia, \emph{{B-decay Anomalies in a Composite
  Leptoquark Model}},
  \href{http://dx.doi.org/10.1140/epjc/s10052-016-4578-7}{\emph{Eur. Phys. J.
  C} {\bf 77} (2017) 8}, [\href{http://arxiv.org/abs/1611.04930}{{\tt
  1611.04930}}].

\bibitem{NP_Cai}
Y.~Cai, J.~Gargalionis, M.~A. Schmidt and R.~R. Volkas, \emph{{Reconsidering
  the One Leptoquark solution: flavor anomalies and neutrino mass}},
  \href{http://dx.doi.org/10.1007/JHEP10(2017)047}{\emph{JHEP} {\bf 10} (2017)
  047}, [\href{http://arxiv.org/abs/1704.05849}{{\tt 1704.05849}}].

\bibitem{NP_Megias}
E.~Megias, M.~Quiros and L.~Salas, \emph{{Lepton-flavor universality violation
  in R$_{K}$ and $ {R}_{D^{{\left(\ast \right)}}} $ from warped space}},
  \href{http://dx.doi.org/10.1007/JHEP07(2017)102}{\emph{JHEP} {\bf 07} (2017)
  102}, [\href{http://arxiv.org/abs/1703.06019}{{\tt 1703.06019}}].

\bibitem{NP_Popov}
O.~Popov, M.~A. Schmidt and G.~White, \emph{{$R_2$ as a single leptoquark
  solution to $R_{D^{(*)}}$ and $R_{K^{(*)}}$}},
  \href{http://dx.doi.org/10.1103/PhysRevD.100.035028}{\emph{Phys. Rev. D} {\bf
  100} (2019) 035028}, [\href{http://arxiv.org/abs/1905.06339}{{\tt
  1905.06339}}].

\bibitem{NP_Angelescu}
A.~Angelescu, D.~Be\v{c}irevi\'c, D.~Faroughy and O.~Sumensari, \emph{{Closing
  the window on single leptoquark solutions to the $B$-physics anomalies}},
  \href{http://dx.doi.org/10.1007/JHEP10(2018)183}{\emph{JHEP} {\bf 10} (2018)
  183}, [\href{http://arxiv.org/abs/1808.08179}{{\tt 1808.08179}}].

\bibitem{NP_Cornella}
C.~Cornella, J.~Fuentes-Martin and G.~Isidori, \emph{{Revisiting the vector
  leptoquark explanation of the B-physics anomalies}},
  \href{http://dx.doi.org/10.1007/JHEP07(2019)168}{\emph{JHEP} {\bf 07} (2019)
  168}, [\href{http://arxiv.org/abs/1903.11517}{{\tt 1903.11517}}].

\bibitem{NP_Marzocca}
D.~Marzocca, \emph{{Addressing the B-physics anomalies in a fundamental
  Composite Higgs Model}},
  \href{http://dx.doi.org/10.1007/JHEP07(2018)121}{\emph{JHEP} {\bf 07} (2018)
  121}, [\href{http://arxiv.org/abs/1803.10972}{{\tt 1803.10972}}].

\bibitem{Isidori}
D.~Buttazzo, A.~Greljo, G.~Isidori and D.~Marzocca, \emph{{B-physics anomalies:
  a guide to combined explanations}},
  \href{http://dx.doi.org/10.1007/JHEP11(2017)044}{\emph{JHEP} {\bf 11} (2017)
  044}, [\href{http://arxiv.org/abs/1706.07808}{{\tt 1706.07808}}].

\bibitem{Crivellin}
A.~Crivellin, D.~M\"uller and F.~Saturnino, \emph{{Flavor Phenomenology of the
  Leptoquark Singlet-Triplet Model}},
  \href{http://dx.doi.org/10.1007/JHEP06(2020)020}{\emph{JHEP} {\bf 06} (2020)
  020}, [\href{http://arxiv.org/abs/1912.04224}{{\tt 1912.04224}}].

\bibitem{NP_Becirevic2}
D.~Be\v{c}irevi\'c, I.~Dor\v{s}ner, S.~Fajfer, N.~Ko\v{s}nik, D.~A. Faroughy
  and O.~Sumensari, \emph{{Scalar leptoquarks from grand unified theories to
  accommodate the $B$-physics anomalies}},
  \href{http://dx.doi.org/10.1103/PhysRevD.98.055003}{\emph{Phys. Rev. D} {\bf
  98} (2018) 055003}, [\href{http://arxiv.org/abs/1806.05689}{{\tt
  1806.05689}}].

\bibitem{Bhupal_1}
K.~Babu, P.~B. Dev, S.~Jana and A.~Thapa, \emph{{Unified Framework for
  $B$-Anomalies, Muon $g-2$, and Neutrino Masses}},
  \href{http://arxiv.org/abs/2009.01771}{{\tt 2009.01771}}.

\bibitem{Bhupal_2}
W.~Altmannshofer, P.~B. Dev, A.~Soni and Y.~Sui, \emph{{Addressing
  R$_{D^{(*)}}$, R$_{K^{(*)}}$, muon $g-2$ and ANITA anomalies in a minimal
  $R$-parity violating supersymmetric framework}},
  \href{http://dx.doi.org/10.1103/PhysRevD.102.015031}{\emph{Phys. Rev. D} {\bf
  102} (2020) 015031}, [\href{http://arxiv.org/abs/2002.12910}{{\tt
  2002.12910}}].

\bibitem{GUT_Fajfer}
I.~Dor\v{s}ner, S.~Fajfer, D.~A. Faroughy and N.~Ko\v{s}nik, \emph{{The role of
  the $S_3$ GUT leptoquark in flavor universality and collider searches}},
  \href{http://dx.doi.org/10.1007/JHEP10(2017)188}{\emph{JHEP} {\bf 10} (2017)
  188}, [\href{http://arxiv.org/abs/1706.07779}{{\tt 1706.07779}}].

\bibitem{Stangl}
J.~Fuentes-Martin and P.~Stangl, \emph{{Third-family quark-lepton unification
  with a fundamental composite Higgs}},
  \href{http://dx.doi.org/10.1016/j.physletb.2020.135953}{\emph{Phys. Lett. B}
  {\bf 811} (2020) 135953}, [\href{http://arxiv.org/abs/2004.11376}{{\tt
  2004.11376}}].

\bibitem{Megias}
E.~Megias, M.~Quiros and L.~Salas, \emph{{Lepton-flavor universality violation
  in R$_{K}$ and $ {R}_{D^{{\left(\ast \right)}}} $ from warped space}},
  \href{http://dx.doi.org/10.1007/JHEP07(2017)102}{\emph{JHEP} {\bf 07} (2017)
  102}, [\href{http://arxiv.org/abs/1703.06019}{{\tt 1703.06019}}].

\bibitem{LQreview}
I.~Dor\v{s}ner, S.~Fajfer, A.~Greljo, J.~Kamenik and N.~Ko\v{s}nik,
  \emph{{Physics of leptoquarks in precision experiments and at particle
  colliders}},
  \href{http://dx.doi.org/10.1016/j.physrep.2016.06.001}{\emph{Phys. Rept.}
  {\bf 641} (2016) 1--68}, [\href{http://arxiv.org/abs/1603.04993}{{\tt
  1603.04993}}].

\bibitem{us}
L.~Da~Rold and F.~Lamagna, \emph{{A vector leptoquark for the B-physics
  anomalies from a composite GUT}},
  \href{http://dx.doi.org/10.1007/JHEP12(2019)112}{\emph{JHEP} {\bf 12} (2019)
  112}, [\href{http://arxiv.org/abs/1906.11666}{{\tt 1906.11666}}].

\bibitem{1412.1791}
B.~Gripaios, M.~Nardecchia and S.~Renner, \emph{{Composite leptoquarks and
  anomalies in $B$-meson decays}},
  \href{http://dx.doi.org/10.1007/JHEP05(2015)006}{\emph{JHEP} {\bf 05} (2015)
  006}, [\href{http://arxiv.org/abs/1412.1791}{{\tt 1412.1791}}].

\bibitem{DaRold:2018moy}
L.~Da~Rold and F.~Lamagna, \emph{{Composite Higgs and leptoquarks from a simple
  group}}, \href{http://dx.doi.org/10.1007/JHEP03(2019)135}{\emph{JHEP} {\bf
  03} (2019) 135}, [\href{http://arxiv.org/abs/1812.08678}{{\tt 1812.08678}}].

\bibitem{Agashe-2004}
K.~Agashe, R.~Contino and A.~Pomarol, \emph{{The Minimal composite Higgs
  model}}, \href{http://dx.doi.org/10.1016/j.nuclphysb.2005.04.035}{\emph{Nucl.
  Phys. B} {\bf 719} (2005) 165--187},
  [\href{http://arxiv.org/abs/hep-ph/0412089}{{\tt hep-ph/0412089}}].

\bibitem{Contino:2004vy}
R.~Contino and A.~Pomarol, \emph{{Holography for fermions}},
  \href{http://dx.doi.org/10.1088/1126-6708/2004/11/058}{\emph{JHEP} {\bf 11}
  (2004) 058}, [\href{http://arxiv.org/abs/hep-th/0406257}{{\tt
  hep-th/0406257}}].

\bibitem{Kaplan:1991dc}
D.~B. Kaplan, \emph{{Flavor at SSC energies: A New mechanism for dynamically
  generated fermion masses}},
  \href{http://dx.doi.org/10.1016/S0550-3213(05)80021-5}{\emph{Nucl. Phys. B}
  {\bf 365} (1991) 259--278}.

\bibitem{Panico:2015jxa}
G.~Panico and A.~Wulzer, \emph{{The Composite Nambu-Goldstone Higgs}},
  vol.~913.
\newblock Springer, 2016,
  \href{http://dx.doi.org/10.1007/978-3-319-22617-0}{10.1007/978-3-319-22617-0}.

\bibitem{2-site-sundrum}
R.~Contino, T.~Kramer, M.~Son and R.~Sundrum, \emph{{Warped/composite
  phenomenology simplified}},
  \href{http://dx.doi.org/10.1088/1126-6708/2007/05/074}{\emph{JHEP} {\bf 05}
  (2007) 074}, [\href{http://arxiv.org/abs/hep-ph/0612180}{{\tt
  hep-ph/0612180}}].

\bibitem{1410.1100}
I.~Phillips, D.G. et~al., \emph{{Neutron-Antineutron Oscillations: Theoretical
  Status and Experimental Prospects}},
  \href{http://dx.doi.org/10.1016/j.physrep.2015.11.001}{\emph{Phys. Rept.}
  {\bf 612} (2016) 1--45}, [\href{http://arxiv.org/abs/1410.1100}{{\tt
  1410.1100}}].

\bibitem{Csaki}
C.~Csaki, A.~Falkowski and A.~Weiler, \emph{{The Flavor of the Composite
  Pseudo-Goldstone Higgs}},
  \href{http://dx.doi.org/10.1088/1126-6708/2008/09/008}{\emph{JHEP} {\bf 09}
  (2008) 008}, [\href{http://arxiv.org/abs/0804.1954}{{\tt 0804.1954}}].

\bibitem{Redi}
M.~Redi and A.~Weiler, \emph{{Flavor and CP Invariant Composite Higgs Models}},
  \href{http://dx.doi.org/10.1007/JHEP11(2011)108}{\emph{JHEP} {\bf 11} (2011)
  108}, [\href{http://arxiv.org/abs/1106.6357}{{\tt 1106.6357}}].

\bibitem{Redi-Weiler}
M.~Redi and A.~Weiler, \emph{{Flavor and CP Invariant Composite Higgs Models}},
  \href{http://dx.doi.org/10.1007/JHEP11(2011)108}{\emph{JHEP} {\bf 11} (2011)
  108}, [\href{http://arxiv.org/abs/1106.6357}{{\tt 1106.6357}}].

\bibitem{Panico-Pomarol}
G.~Panico and A.~Pomarol, \emph{{Flavor hierarchies from dynamical scales}},
  \href{http://dx.doi.org/10.1007/JHEP07(2016)097}{\emph{JHEP} {\bf 07} (2016)
  097}, [\href{http://arxiv.org/abs/1603.06609}{{\tt 1603.06609}}].

\bibitem{1807.04279}
M.~Frigerio, M.~Nardecchia, J.~Serra and L.~Vecchi, \emph{{The Bearable
  Compositeness of Leptons}},
  \href{http://dx.doi.org/10.1007/JHEP10(2018)017}{\emph{JHEP} {\bf 10} (2018)
  017}, [\href{http://arxiv.org/abs/1807.04279}{{\tt 1807.04279}}].

\bibitem{1708.08515}
L.~Da~Rold, \emph{{Anarchy with linear and bilinear interactions}},
  \href{http://dx.doi.org/10.1007/JHEP10(2017)120}{\emph{JHEP} {\bf 10} (2017)
  120}, [\href{http://arxiv.org/abs/1708.08515}{{\tt 1708.08515}}].

\bibitem{Callan:1969sn}
C.~G. Callan, Jr., S.~R. Coleman, J.~Wess and B.~Zumino, \emph{{Structure of
  phenomenological Lagrangians. 2.}},
  \href{http://dx.doi.org/10.1103/PhysRev.177.2247}{\emph{Phys. Rev.} {\bf 177}
  (1969) 2247--2250}.

\bibitem{Carena-2014}
M.~Carena, L.~Da~Rold and E.~Pont\'on, \emph{{Minimal Composite Higgs Models at
  the LHC}}, \href{http://dx.doi.org/10.1007/JHEP06(2014)159}{\emph{JHEP} {\bf
  06} (2014) 159}, [\href{http://arxiv.org/abs/1402.2987}{{\tt 1402.2987}}].

\bibitem{1210.7114}
G.~Panico, M.~Redi, A.~Tesi and A.~Wulzer, \emph{{On the Tuning and the Mass of
  the Composite Higgs}},
  \href{http://dx.doi.org/10.1007/JHEP03(2013)051}{\emph{JHEP} {\bf 03} (2013)
  051}, [\href{http://arxiv.org/abs/1210.7114}{{\tt 1210.7114}}].

\bibitem{0612048}
R.~Contino, L.~Da~Rold and A.~Pomarol, \emph{{Light custodians in natural
  composite Higgs models}},
  \href{http://dx.doi.org/10.1103/PhysRevD.75.055014}{\emph{Phys. Rev. D} {\bf
  75} (2007) 055014}, [\href{http://arxiv.org/abs/hep-ph/0612048}{{\tt
  hep-ph/0612048}}].

\bibitem{DeCurtis}
S.~De~Curtis, M.~Redi and A.~Tesi, \emph{{The 4D Composite Higgs}},
  \href{http://dx.doi.org/10.1007/JHEP04(2012)042}{\emph{JHEP} {\bf 04} (2012)
  042}, [\href{http://arxiv.org/abs/1110.1613}{{\tt 1110.1613}}].

\bibitem{hflavURL}
HFLAV, \emph{Average for $R(D)$ and $R(D^*)$ for Spring 2019}.
\newblock
  https://hflav-eos.web.cern.ch/hflav-eos/semi/spring19/html/RDsDsstar/RDRDs.html.

\bibitem{dC9}
B.~Capdevila, A.~Crivellin, S.~Descotes-Genon, J.~Matias and J.~Virto,
  \emph{{Patterns of New Physics in $b\to s\ell^+\ell^-$ transitions in the
  light of recent data}},
  \href{http://dx.doi.org/10.1007/JHEP01(2018)093}{\emph{JHEP} {\bf 01} (2018)
  093}, [\href{http://arxiv.org/abs/1704.05340}{{\tt 1704.05340}}].

\bibitem{PDG}
{\scshape Particle Data Group} collaboration, P.~Zyla et~al., \emph{{Review of
  Particle Physics}}, \href{http://dx.doi.org/10.1093/ptep/ptaa104}{\emph{PTEP}
  {\bf 2020} (2020) 083C01}.

\bibitem{1702.03224}
{\scshape Belle} collaboration, J.~Grygier et~al., \emph{{Search for
  $\boldsymbol{B\to h\nu\bar{\nu}}$ decays with semileptonic tagging at
  Belle}}, \href{http://dx.doi.org/10.1103/PhysRevD.96.091101}{\emph{Phys. Rev.
  D} {\bf 96} (2017) 091101}, [\href{http://arxiv.org/abs/1702.03224}{{\tt
  1702.03224}}]. [Addendum: Phys.Rev.D 97, 099902 (2018)].

\bibitem{1504.07928}
A.~Crivellin, L.~Hofer, J.~Matias, U.~Nierste, S.~Pokorski and J.~Rosiek,
  \emph{{Lepton-flavour violating $B$ decays in generic $Z'$ models}},
  \href{http://dx.doi.org/10.1103/PhysRevD.92.054013}{\emph{Phys. Rev. D} {\bf
  92} (2015) 054013}, [\href{http://arxiv.org/abs/1504.07928}{{\tt
  1504.07928}}].

\bibitem{1609.09078}
B.~Bhattacharya, A.~Datta, J.-P. Gu\'evin, D.~London and R.~Watanabe,
  \emph{{Simultaneous Explanation of the $R_K$ and $R_{D^{(*)}}$ Puzzles: a
  Model Analysis}},
  \href{http://dx.doi.org/10.1007/JHEP01(2017)015}{\emph{JHEP} {\bf 01} (2017)
  015}, [\href{http://arxiv.org/abs/1609.09078}{{\tt 1609.09078}}].

\bibitem{1811.09603}
M.~Blanke, A.~Crivellin, S.~de~Boer, T.~Kitahara, M.~Moscati, U.~Nierste
  et~al., \emph{{Impact of polarization observables and $ B_c\to \tau \nu$ on
  new physics explanations of the $b\to c \tau \nu$ anomaly}},
  \href{http://dx.doi.org/10.1103/PhysRevD.99.075006}{\emph{Phys. Rev. D} {\bf
  99} (2019) 075006}, [\href{http://arxiv.org/abs/1811.09603}{{\tt
  1811.09603}}].

\bibitem{1811.08899}
S.~Iguro, T.~Kitahara, Y.~Omura, R.~Watanabe and K.~Yamamoto, \emph{{D$^{*}$
  polarization vs. $ {R}_{D^{\left(\ast \right)}} $ anomalies in the leptoquark
  models}}, \href{http://dx.doi.org/10.1007/JHEP02(2019)194}{\emph{JHEP} {\bf
  02} (2019) 194}, [\href{http://arxiv.org/abs/1811.08899}{{\tt 1811.08899}}].

\bibitem{1706.00410}
M.~Gonz\'alez-Alonso, J.~Martin~Camalich and K.~Mimouni,
  \emph{{Renormalization-group evolution of new physics contributions to
  (semi)leptonic meson decays}},
  \href{http://dx.doi.org/10.1016/j.physletb.2017.07.003}{\emph{Phys. Lett. B}
  {\bf 772} (2017) 777--785}, [\href{http://arxiv.org/abs/1706.00410}{{\tt
  1706.00410}}].

\bibitem{1008.1593}
A.~Lenz, U.~Nierste, J.~Charles, S.~Descotes-Genon, A.~Jantsch, C.~Kaufhold
  et~al., \emph{{Anatomy of New Physics in $B - \bar{B}$ mixing}},
  \href{http://dx.doi.org/10.1103/PhysRevD.83.036004}{\emph{Phys. Rev. D} {\bf
  83} (2011) 036004}, [\href{http://arxiv.org/abs/1008.1593}{{\tt 1008.1593}}].

\bibitem{1302.0661}
G.~Isidori, \emph{{Flavor physics and CP violation}},  in \emph{{2012 European
  School of High-Energy Physics}}, pp.~69--105, 2014.
\newblock \href{http://arxiv.org/abs/1302.0661}{{\tt 1302.0661}}.
\newblock \href{http://dx.doi.org/10.5170/CERN-2014-008.69}{DOI}.

\bibitem{Arnan}
P.~Arnan, D.~Becirevic, F.~Mescia and O.~Sumensari, \emph{{Probing low energy
  scalar leptoquarks by the leptonic $W$ and $Z$ couplings}},
  \href{http://dx.doi.org/10.1007/JHEP02(2019)109}{\emph{JHEP} {\bf 02} (2019)
  109}, [\href{http://arxiv.org/abs/1901.06315}{{\tt 1901.06315}}].

\bibitem{hep-ex/0509008}
{\scshape ALEPH, DELPHI, L3, OPAL, SLD, LEP Electroweak Working Group, SLD
  Electroweak Group, SLD Heavy Flavour Group} collaboration, S.~Schael et~al.,
  \emph{{Precision electroweak measurements on the $Z$ resonance}},
  \href{http://dx.doi.org/10.1016/j.physrep.2005.12.006}{\emph{Phys. Rept.}
  {\bf 427} (2006) 257--454}, [\href{http://arxiv.org/abs/hep-ex/0509008}{{\tt
  hep-ex/0509008}}].

\bibitem{Janot}
P.~Janot and S.~Jadach, \emph{{Improved Bhabha cross section at LEP and the
  number of light neutrino species}},
  \href{http://dx.doi.org/10.1016/j.physletb.2020.135319}{\emph{Phys. Lett. B}
  {\bf 803} (2020) 135319}, [\href{http://arxiv.org/abs/1912.02067}{{\tt
  1912.02067}}].

\bibitem{0908.2381}
{\scshape BaBar} collaboration, B.~Aubert et~al., \emph{{Searches for Lepton
  Flavor Violation in the Decays tau+- ---\ensuremath{>} e+- gamma and tau+-
  ---\ensuremath{>} mu+- gamma}},
  \href{http://dx.doi.org/10.1103/PhysRevLett.104.021802}{\emph{Phys. Rev.
  Lett.} {\bf 104} (2010) 021802}, [\href{http://arxiv.org/abs/0908.2381}{{\tt
  0908.2381}}].

\bibitem{MEG}
{\scshape MEG} collaboration, A.~M. Baldini et~al., \emph{{Search for the
  lepton flavour violating decay $\mu ^+ \rightarrow \mathrm {e}^+ \gamma $
  with the full dataset of the MEG experiment}},
  \href{http://dx.doi.org/10.1140/epjc/s10052-016-4271-x}{\emph{Eur. Phys. J.
  C} {\bf 76} (2016) 434}, [\href{http://arxiv.org/abs/1605.05081}{{\tt
  1605.05081}}].

\bibitem{Matsedonskyi}
O.~Matsedonskyi, \emph{{On Flavour and Naturalness of Composite Higgs Models}},
  \href{http://dx.doi.org/10.1007/JHEP02(2015)154}{\emph{JHEP} {\bf 02} (2015)
  154}, [\href{http://arxiv.org/abs/1411.4638}{{\tt 1411.4638}}].

\bibitem{1705.00929}
F.~Feruglio, P.~Paradisi and A.~Pattori, \emph{{On the Importance of
  Electroweak Corrections for B Anomalies}},
  \href{http://dx.doi.org/10.1007/JHEP09(2017)061}{\emph{JHEP} {\bf 09} (2017)
  061}, [\href{http://arxiv.org/abs/1705.00929}{{\tt 1705.00929}}].

\bibitem{1808.10567}
{\scshape Belle-II} collaboration, W.~Altmannshofer et~al., \emph{{The Belle II
  Physics Book}}, \href{http://dx.doi.org/10.1093/ptep/ptz106}{\emph{PTEP} {\bf
  2019} (2019) 123C01}, [\href{http://arxiv.org/abs/1808.10567}{{\tt
  1808.10567}}]. [Erratum: PTEP 2020, 029201 (2020)].

\bibitem{mu3e}
U.~Bellgardt, G.~Otter, R.~Eichler, L.~Felawka, C.~Niebuhr, H.~Walter et~al.,
  \emph{Search for the decay $\mu+ \to e+ e+ e-$},
  \href{http://dx.doi.org/https://doi.org/10.1016/0550-3213(88)90462-2}{\emph{Nuclear
  Physics B} {\bf 299} (1988) 1--6}.

\bibitem{pich}
A.~Pich, \emph{{Precision Tau Physics}},
  \href{http://dx.doi.org/10.1016/j.ppnp.2013.11.002}{\emph{Prog. Part. Nucl.
  Phys.} {\bf 75} (2014) 41--85}, [\href{http://arxiv.org/abs/1310.7922}{{\tt
  1310.7922}}].

\bibitem{1808.02063}
E.~Alvarez, L.~Da~Rold, A.~Juste, M.~Szewc and T.~Vazquez~Schroeder, \emph{{A
  composite pNGB leptoquark at the LHC}},
  \href{http://dx.doi.org/10.1007/JHEP12(2018)027}{\emph{JHEP} {\bf 12} (2018)
  027}, [\href{http://arxiv.org/abs/1808.02063}{{\tt 1808.02063}}].

\bibitem{2101.11582}
{\scshape ATLAS} collaboration, G.~Aad et~al., \emph{{Search for pair
  production of third-generation scalar leptoquarks decaying into a top quark
  and a $\tau$-lepton in $pp$ collisions at $\sqrt{s}=13$ TeV with the ATLAS
  detector}},  \href{http://arxiv.org/abs/2101.11582}{{\tt 2101.11582}}.

\bibitem{2101.12527}
{\scshape ATLAS} collaboration, G.~Aad et~al., \emph{{Search for new phenomena
  in final states with $b$-jets and missing transverse momentum in
  $\sqrt{s}=13$ TeV $pp$ collisions with the ATLAS detector}},
  \href{http://arxiv.org/abs/2101.12527}{{\tt 2101.12527}}.

\bibitem{2012.04178}
{\scshape CMS} collaboration, A.~M. Sirunyan et~al., \emph{{Search for singly
  and pair-produced leptoquarks coupling to third-generation fermions in
  proton-proton collisions at $\sqrt{s} =$ 13 TeV}},
  \href{http://arxiv.org/abs/2012.04178}{{\tt 2012.04178}}.

\bibitem{Raby}
S.~Raby, \emph{{Supersymmetric Grand Unified Theories}: {From Quarks to Strings
  via SUSY GUTs}}, vol.~939.
\newblock Springer, 2017,
  \href{http://dx.doi.org/10.1007/978-3-319-55255-2}{10.1007/978-3-319-55255-2}.

\end{thebibliography}\endgroup

\end{document}